\documentclass[longauth]{aa}

\newcommand{\bbv}[1]{\textbf{\textcolor{cyan}{(BV: #1)}}}

\usepackage{color}
\usepackage{graphicx}
\usepackage{natbib}
\usepackage{enumitem}
\usepackage{txfonts}
\usepackage{multirow}
\usepackage{rotating}
\usepackage{longtable}
\usepackage{amssymb}
\usepackage{verbatim}
\usepackage{caption}
\usepackage{subcaption}

\titlerunning{Analysis on XLSSsC N01 supercluster}
\authorrunning{V. Guglielmo et al.}
\begin{document}

\title{The XXL Survey: XXX. Characterisation of the XLSSsC N01 supercluster and analysis of the galaxy stellar populations}

\author{V. Guglielmo \inst{1,2,3} \and B. M. Poggianti \inst{1} \and B. Vulcani \inst{4,1} \and A. Moretti \inst{1} \and J. Fritz \inst{5} \and F. Gastaldello \inst{6} \and C. Adami \inst{2} \and C. A. Caretta \inst{7}  \and J.Willis \inst{12} \and E. Koulouridis \inst{8,9} \and M. E. Ramos Ceja \inst{11} \and P. Giles \inst{12} \and I. Baldry \inst{14} \and M. Birkinshaw \inst{12} \and A. Bongiorno \inst{15} \and M. Brown \inst{16} \and L. Chiappetti \inst{6} \and S. Driver \inst{17,18} \and A. Elyiv \inst{19,20} \and A. Evrard \inst{21} \and M. Grootes \inst{22} \and L. Guennou \inst{23} \and A. Hopkins \inst{24} \and C. Horellou \inst{25} \and A. Iovino \inst{26} \and S. Maurogordato \inst{27} \and M. Owers \inst{28} \and F. Pacaud \inst{11} \and S. Paltani \inst{29} \and M. Pierre \inst{8,9} \and M. Plionis \inst{30,31} \and T. Ponman  \inst{32} \and A. Robotham \inst{17}  \and T. Sadibekova \inst{33} \and V. Smol\v ci\'c \inst{34} \and R. Tuffs \inst{22} \and C. Vignali \inst{10,19}
} 

\institute{INAF-Osservatorio astronomico di Padova, Vicolo Osservatorio 5, IT-35122 Padova, Italy\\
              \email{valentina.guglielmo@oapd.inaf.it}
              \and
             Aix Marseille Universit\'e, CNRS, LAM (Laboratoire d'Astrophysique de Marseille) UMR 7326, F-13388, Marseille, France
             \and Max-Planck-Institut f{\"u}r Extraterrestriche Physik, Giessenbachstrasse, D-85748 Garching, Germany
             \and School of Physics, University of Melbourne, VIC 3010, Australia
           \and Instituto de Radioastronom{\`i}a y Astrof{\`i}sica, UNAM, Campus Morelia, A.P. 3-72, C.P. 58089, Mexico
           \and INAF-Istituto di Astrofisica Spaziale e Fisica Cosmica Milano, Via Bassini 15, I-20133 Milan, Italy
           \and Departamento de Astronom{\`i}a, DCNE-CGT, Universidad de Guanajuato; Callej\'on de Jalisco, S/N, Col. Valenciana, 36240, Guanajuato, Gto., Mexico
           \and IRFU, CEA, Universit\'e Paris-Saclay, F-91191 Gif-sur-Yvette, France\\
           \and Universit\'e Paris Diderot, AIM, Sorbonne Paris Cit\'e, CEA, CNRS, F-91191
Gif-sur-Yvette, France
			\and INAF-Osservatorio Astronomico di Bologna, via Gobetti 93/3, I-40129 Bologna, Italy
            \and Argelander Institut für Astronomie, Universität Bonn, Auf dem Huegel 71, D-53121 Bonn, Germany
            \and School of Physics, HH Wills Physics Laboratory, Tyndall Avenue, Bristol, GB-BS8 1TL, UK
            \and Department of Physics and Astronomy, University of Victoria, 3800 Finnerty Road, Victoria, BC, Canada    
            \and Astrophysics Research Institute, Liverpool John Moores University, IC2, Liverpool Science Park, 146 Brownlow Hill, Liverpool GB-L3 5RF, UK
           \and INAF-Osservatorio Astronomico di Roma, via Frascati 33, I-00078 Monte Porzio Catone (Rome), Italy
           \and School of Physics and Astronomy, Monash University, Clayton, Victoria AU-3800, Australia 
           \and International Centre for Radio Astronomy Research (ICRAR), The University of Western Australia, M468, 35 Stirling Highway, Crawley, WA AU-6009, Australia
           \and SUPA, School of Physics \& Astronomy, University of St Andrews, North Haugh, St Andrews GB-KY16 9SS, UK
           \and Dipartimento di Fisica e Astronomia, Alma Mater Studiorum – Università di Bologna, via Gobetti 93/3, I-40129 Bologna, Italy
            \and Main Astronomical Observatory, Academy of Sciences of Ukraine, 27 Akademika Zabolotnoho St., UA-03680 Kyiv, Ukraine
            \and Department of Physics and Michigan Center for Theoretical Physics, University of Michigan, Ann Arbor, MI Us-48109, USA
            \and ESA/ESTEC, Noordwijk 2201 AZ, The Netherlands
            \and Astrophysics and Cosmology Research Unit, University of KwaZulu-Natal, ZA-4041 Durban, South Africa
            \and Australian Astronomical Observatory, PO BOX 915, North Ryde, AU-1670, Australia
            \and Chalmers University of Technology, Dept. of Space, Earth and Environment, Onsala Space Observatory, SE-439 92 Onsala, Sweden
            \and INAF-Osservatorio Astronomico di Brera, via Brera, 28, I-20121 Milano, Italy
            \and Laboratoire Lagrange, UMR 7293, Universit\'e de Nice Sophia Antipolis, CNRS, Observatoire de la C\^{o}te d’Azur, FR-06304 Nice, France
            \and Department of Physics and Astronomy, Macquarie University, NSW 2109, Australia and Australian Astronomical Observatory PO Box 915, North Ryde NSW AU-1670, Australia
            \and Department of Astronomy, University of Geneva, ch. d’Ecogia 16, CH-1290 Versoix, Switzerland
            \and Aristotle University of Thessaloniki, Physics Department, GR-54124 Thessaloniki, Greece
            \and IAASARS, National Observatory of Athens, GR-15236 Penteli, Greece
            \and School of Physics and Astronomy, University of Birmingham, Edgbaston, Birmingham GB-B15 2TT,UK
            \and Ulugh Beg Astronomical Institute of Uzbekistan Academy of Science, 33 Astronomicheskaya str., Tashkent, UZ-100052, Uzbekistan
            \and Department of Physics, University of Zagreb, Bijenicka cesta 32, HR-10000 Zagreb, Croatia
            }

\date{Received xxx; accepted yyy}

\abstract
{Superclusters form from the largest enhancements in the primordial density perturbation field and extend for tens of Mpc, tracing the large-scale structure of the Universe. X-ray detections and systematic characterisations of superclusters and the properties of their galaxies have only been possible in the last few years.}
{We characterise XLSSsC N01, a rich supercluster at z$\sim$0.3 detected in the XXL Survey, composed of X-ray clusters of different virial masses and X-ray luminosities. As one of the first studies on this topic, we investigate the stellar populations of galaxies in different environments in the supercluster region.}
{We study a magnitude-limited ($r\leq$20) and a  mass-limited sample ($\log (M_{*}/M_{\odot}) \geq 10.8$) of galaxies in the virialised region and in the outskirts of 11 XLSSsC N01 clusters, in high-density field regions, and in the low-density field.
We compute the stellar population properties of galaxies using spectral energy distribution (SED) and spectral fitting techniques, and study the dependence of star formation rates (SFR), colours,  and stellar ages on environment.}
{For $r\leq$20, the fraction of star-forming/blue galaxies, computed either from the specific-SFR (sSFR) or rest-frame colour, shows depletion within the cluster virial radii, where the number of galaxies with $\rm \log(sSFR/ yr^{-1})>-12$ and with $\rm (g-r)_{restframe}<0.6$ is lower than in the field. For $\log(M_{*}/M_{\odot})\geq 10.8$, no trends with environment emerge, as massive galaxies are mostly already passive in all environments.
No differences among low- and high-density field members and cluster members emerge in the sSFR-mass relation in the mass-complete regime. Finally, the luminosity-weighted age-mass relation of the passive populations within cluster virial radii show signatures of recent environmental quenching.}
{The study of luminous and massive galaxies in this supercluster shows that  while environment has a prominent role in determining the fractions of star-forming/blue galaxies, its effects on the star formation activity in star-forming galaxies are negligible.}


\maketitle

\section{Introduction}

Galaxy stellar population properties, especially  star formation history and colour, depend on the environment in which  galaxies reside \citep{Spitzer1951,Oemler1974,Davis1976,Dressler1980,Blanton2005,Ball2008}.
On the other hand, the galaxy stellar mass also plays a significant role in determining these properties \citep{Scodeggio2002,Kauffmann2003}.

Stellar mass and environment have been found to be the main drivers of galaxy transformations in different regimes. 
Overall, the environment seems to be more relevant for lower mass galaxies, at least as far as quenching is concerned: galaxies in denser environments tend to be redder than galaxies in less dense environments \citep{Haines2007,Cooper2010,Pasquali2010,Peng2010,Peng2012,McGee2011,Sobral2011,Muzzin2012,Smith2012,Wetzel2012,LaBarbera2014,Lin2014,Vulcani2015}. 
By contrast, on average, more massive galaxies have formed their stars and completed their star formation activity at higher redshift than less massive galaxies  \citep[known as the  downsizing effect; ][]{Cowie1996,Gavazzi2006,DeLucia2007,SanchezBlazquez2009}, regardless of environment.

While different methods of estimating stellar masses agree reasonably well within the errors \citep[e.g.][]{Bell2001,Bolzonella2010}, different definitions of environment do not always probe the same scales \citep[e.g.][]{Muldrew2012,Fossati2015}. It is possible to discuss environmental effects in either the global or local context.
Concerning the global environment, galaxies are commonly subdivided into  superclusters, clusters, groups, and field and void populations. These roughly correspond to halos of different mass. The local environment is generally described using estimates of projected local density, which can be calculated following several definitions and methods \citep[e.g.][]{KovaC2010,Cucciati2010,Muldrew2012,Vulcani2012, Darvish2015,Fasano2015}.

Considering the local density, in the local Universe, \citet{Baldry2006} found that the fraction of galaxies in the red sequence is higher in denser environments at any stellar mass in the range $9.0<\log(M_\ast/M_{\odot})<11.0$.
Similar results have been confirmed at higher redshift \citep[e.g.][]{Scoville2007,Cucciati2010,Cucciati2017}, 
where many studies show that all features of the global correlation between galaxy colour and environment measured at $z\sim 0$ (known as galaxy bimodality) are already in place at $z\sim 1$, with blue galaxies on average occupying regions of lower density than red galaxies \citep[e.g.][]{Coil2008,Cooper2006,Cooper2010,Wilman2005,Cucciati2006}. At these redshifts, the inverse of the specific star formation rate (sSFR), i.e. the time for a galaxy to double its stellar mass, is higher in denser environments \citep{Scoville2013}. 
Both in the local Universe and at higher redshift, the stellar mass distribution is also sensitive to the local environment, in the sense that more massive galaxies are preferentially found at higher densities \citep[e.g.][]{Hogg2003,Kauffmann2004,Blanton2005,Bolzonella2010,Cucciati2010,Vulcani2012,Davidzon2016}. 

In contrast to these studies, \citet{Scodeggio2009} showed that while there is evidence for a colour-density relation at fixed luminosity at z $\sim$ 1, at intermediate redshifts and fixed stellar mass  no colour-density relation can be found. At similar redshifts, \citet{Tasca2009} did not find any variation in galaxy morphology (i.e. early- versus late-type) as a function of local galaxy density for $\log(M_\ast/M_{\odot}) > 10.5$.  These works both concluded that the properties (colour and morphology) of massive galaxies are independent of environment.

With regard to  the global environment, star formation quenching seems to be stronger in clusters, which display higher fractions of red early-type and lower fractions of blue late-type galaxies than the field \citep[e.g.][]{Dressler1980,Poggianti1999,Bai2009,Vulcani2013}, suggesting that clusters are extremely effective in cutting off the galaxy's ability to form stars.
In an evolutionary scenario,  \cite{Poggianti2006} and \cite{Iovino2010}, showed that galaxy clusters and groups have seen an evolution in their star-forming galaxy fractions that is stronger than in the field, and that the evolution from blue star-forming to red passive types is faster in dense environments and massive halos. This scenario implies that the fraction of red passive galaxies in clusters increases earlier than in the field, supporting again the environmental quenching.

Focusing on the star-forming population, no consensus regarding the properties of active galaxies in the cluster population relative to those in the field has been reached. Some studies have suggested that the difference in the star formation activity between the field and clusters is primarily driven by the relative red fraction instead of the properties of the star-forming population \citep{Muzzin2012,Koyama2013,Lin2014,Jian2018}. For example, \cite{Lin2014} and  \cite{Jian2018} have found that the sSFRs of star-forming galaxies in clusters are only moderately lower than those in the field (< 0.2--0.3 dex) and that the difference becomes insignificant at group scale.
In contrast, other works (e.g. \citealt{Patel2009,Vulcani2010} at intermediate redshift and \citealt{Haines2015,Paccagnella2016} at low-redshift) identified the presence of a population with reduced star formation rate in clusters with respect to the field, suggesting that both the relative numbers of blue star-forming and red passive galaxies and the characteristics of the star-forming population change with environment.

Regarding the global environment, in addition to studies relating the cluster and field populations, there is an increasing focus on even larger structures,  superclusters. Superclusters are defined as the most extended density enhancements formed from primordial perturbations on scales of about 100 $h^{-1}$ Mpc ($H_0 = 100 h\, km\, s^{-1} Mpc^{-1}$) \citep{Bahcall1984}, and presenting a variety of characterising properties such as morphology, luminosity, and richness \citep{Einasto2011a,Einasto2011b}.

Studying the properties of superclusters helps us to understand the formation, evolution, and properties of the large-scale structure of the Universe \citep[and references therein]{Hoffman2007,Araya-Melo2009,Bond2010}, ranging from rich, large superclusters containing many massive clusters and extending over 10--20 Mpc down to less massive structures containing groups and poor clusters of the order of $10^{13}-10^{14} M_\odot$ each \citep[e.g.][and references therein]{Einasto2011a}.

The majority of supercluster catalogues  in the literature are based on optical data, and have been constructed using the friend-of-friend method  or using a smoothed density field of galaxies. Only in the last few years have searches for superclusters based purely on X-ray detection been pursued out to z $\leq$ 0.4 \citep{Chon2013}.

Even  rarer than systematic studies on the characterisation of the properties of supercluster structures as a whole (\citealt{Einasto2011a} in the SDSS survey; \citealt{Verdugo2012,Geach2011,Schirmer2011,Lubin2009,Kartaltepe2008,Tanaka2007} at z$\geq$0.4) are studies of stellar population properties of galaxies that inhabit such environments.
A connection between supercluster environment and star formation has started to emerge \citep{Lietzen2012,Costa-Duarte2013,Cohen2017,Luparello2013}.
At low redshift, using SDSS data, \citet{Lietzen2012} studied the spectral properties of galaxies exploiting the whole range of large-scale environments from voids to superclusters, and also identified the group-scale environment and the group richness. They found that within superclusters the fraction of passive galaxies increases independently from the morphology. Furthermore, the fraction of passive galaxies increases in rich groups when they are located within superclusters, where  equally rich groups are also more luminous than their counterparts in voids.
Recently, \cite{Cohen2017} have analysed the relationship among  star formation, the amount of cluster substructure, and supercluster environment in a sample of 107 nearby galaxy clusters using data from SDSS, finding a significant inverse correlation between the density of the supercluster environment and the fraction of star-forming galaxies within clusters. 
Furthermore, using low redshift data from the SDSS, \cite{Luparello2013} showed that galaxies in groups residing in superclusters have greater stellar mass content and longer timescales  for star formation than the typical values found in groups outside superclusters, regardless of distance to the group centre.  They concluded that, according to the assembly bias scenario, groups in superclusters formed earlier than elsewhere.

A few isolated, higher redshift superclusters are known \citep[e.g.][]{Gal2004,Swinbank2007,Guzzo2007,Gilbank2008,Iovino2016,Kim2016}. \cite{Pompei2016} (hereafter XXL Paper VII) provide one of the first examples of such a supercluster found in a homogeneous X-ray sample using XXL survey data.
The XXL survey \citep[hereafter XXL Paper I]{Pierre2016}, is an extension of the XMM-LSS 11 $\rm deg^2$ survey \citep{Pierre2004}, consisting of 622 XMM pointings covering a total area of $\sim 50 \, {\rm deg^2}$. The survey reaches a sensitivity of $\rm \sim 5 \times 10^{-15} erg \, s^{-1} \, cm^{-2}$ in the [0.5--2] keV band for point sources.
Within the astrophysical context outlined above, XXL provides an unprecedented volume between 0.1 < z < 1 within which is it possible to study the nature and evolutionary properties of groups, clusters, and superclusters of galaxies.
After the extended ROSAT-ESO Flux-Limited X-ray Galaxy Cluster Survey (REFLEX II), the XXL survey is the second to have detected several superclusters of galaxies beyond $z = 0.4$. As already highlighted in \citet[][hereafter XXL Paper II]{Pacaud2016} and  in Adami et. al (submitted, hereafter XXL Paper XX), the selection method used for XXL superclusters has the advantages of relying only on galaxy structures showing clear evidence of a deep potential well and of extending the volume used for such  studies (to median $z \geq 0.3$).

The present work is focused on XLSSsC N01, a supercluster located in the XXL-North (XXL-N) field centred at RA=36.952$^\circ$, Dec=-4775$^\circ$, and with centroid redshift $z$=0.2956 (XXL Paper XX). This supercluster is the best candidate for environmental studies on galaxies since it is the richest in the XXL-N field (14 groups and clusters; hereafter simply clusters), and because it is located in a region of the sky with highly complete spectroscopic and photometric data. This analysis is a first attempt to directly study the impact of large scale environment, i.e. on clusters which are considered all part of the same superstructure,   on the star formation activity and stellar population properties of galaxies.

The aim of this work is to present the XLSSsC N01 supercluster, to characterise the clusters it is composed of, and to investigate the stellar population properties of galaxies classified as members or belonging to the surrounding high-density and low-density fields.
The paper is organised as follows: in Section 2 we present the datasets and the tools that we used to compute galaxy stellar population properties; in Section 3 we characterise different environments in the region of the XLSSsC N01 supercluster; in Section 4 we explore the dependence of the stellar population properties on environment; and finally in Section 5 we summarise this work and  discuss our results.

Throughout the paper we assume $H_0 = 69.3 \, km \, s^{-1} \,Mpc^{-1}, \, \Omega_M = 0.29, \, \Omega_{\Lambda} = 0.71$ \citep[][Planck13+Alens]{Planck2014}. We adopt a \cite{Chabrier2003} initial mass function (IMF) in the mass range $0.1-100 M_{\odot}$.

\section{Datasets and tools}

\subsection{Catalogue of the structures}
We base our analysis on X-ray selected structures identified within the XXL survey.
The observing strategy and science goals of the survey are described in XXL Paper I, while the source selection is presented in XXL Paper II.
Briefly, the X-ray images were processed with the \textsc{Xamin} pipeline \citep{Pacaud2006}, which produces lists of detections of cluster candidates grouped into detection classes on the basis of the level of contamination from point-sources. Class 1 (C1) includes the highest surface brightness extended sources, with no contamination from point sources; class 2 (C2) includes extended sources fainter than those classified as C1, with a 50\% contamination rate before visual inspection. Contaminating sources include saturated point sources, unresolved pairs, and sources strongly masked by CCD gaps, for which not enough photons were available to permit reliable source characterisation; the third class, C3, corresponds to optical clusters associated with some X-ray emission, too weak to be characterised; initially, most of the C3 objects were not detected in the X-ray waveband and are located within the XMM-LSS subregion, and their selection function is therefore undefined. These objects are not included in the scientific analysis of this work. 365 extended sources were identified, 207  of which ($\sim 56\%$) are classified as C1, 119 ($\sim 32\%$) as C2, and the remaining 39 ($\sim 11\%$) are C3. 

XXL Paper XX presents the spectroscopic confirmation of cluster candidates (see also Guglielmo et al. 2017, hereafter XXL Paper XXII),  based on an iterative semi-automatic process, very similar to that already used for the XMM-LSS survey (e.g. \citealt{Adami2011}). 
XXL Paper XX also releases the final catalogues of 365 spectroscopically confirmed clusters. The 212 clusters brighter than $\sim$1.3 $\times$10$^{-14}$  $\rm erg \, s^{-1} \, cm^{-2}$  underwent dedicated X-ray luminosity and temperature measurements. To have homogeneous estimates for the complete sample, scaling relations based on the $r=300$ kpc count-rates were applied (see XXL Paper XX).
The starting point for deriving scaling relations is the measure of XMM count rates in the $0.5-2$ keV band extracted within 300 kpc of the cluster centres, and from which the temperature of the gas ($T_{300kpc,scal}$) can be derived. The procedure then iterates on the temperature to recover the $M_{500,scal}$ (using the $M-T$ relation derived for the sample XXL+COSMOS+CCCP in Table 2 of paper IV) and $r_{500,scal}$\footnote{$r_{500,scal}$ is defined as the radius of the sphere inside which the mean density is 500 times the critical density $\rho_c$ of the Universe at the cluster redshift; $M_{500,scal}$ is then defined as $4/3 \pi 500 \rho_c r_{500,scal}^3$} and a luminosity $L^{XXL}_{500,scal}$ (using the best-fit results for the relation $L_{XXL}-T$, with the `XXL fit' in Table 2 of paper III) that was integrated up to $r_{500}$ by adopting a $\beta$-model with parameters $(r_c, \beta) = (0.15 r_{500}, 2/3)$.
This method provides estimates and relative errors propagated from the best-fit results of the X-ray temperature ($T_{300kpc,scal}$), $r_{500,scal}$, $M_{500,scal}$, and the luminosity in the 0.5--2.0 keV range ($L^{XXL}_{500,scal}$). 
The values used in the current paper are extracted from XXL Paper XX, where the authors also performed  a comparison between the measured cluster temperatures and those obtained from the scaling relations.
Furthermore, after deriving the virial mass $M_{200}$ from $M_{500,scal}$ using the recipe given in \cite{Balogh2006}, we compute velocity dispersions ($\sigma_{200}$) through the relation given in \cite{Poggianti2006}, based on the virial theorem:
\begin{equation}
\sigma_{200} = 1000 \, {\rm km \, s^{-1}} \cdot \left(\frac{M_{200}}{1.2 \cdot 10^{15} M_\odot} \cdot \sqrt{\Omega_{\Lambda} + \Omega_0(1+z)^3} \cdot h\right)^{1/3}
\label{sigma_200}
\end{equation}

Finally, 35 superclusters were identified in the XXL-N and XXL-S fields in the 0.03$\leq$z$\leq$1.0 redshift range by means of a friend-of-friend (FoF) algorithm  characterised by a Voronoi tesselation technique; the complete list is available in Table 9 of XXL Paper XX.
The aim is to look for physical associations between individual clusters of galaxies, and call `superclusters' the associations of at least three clusters.
First, a classical FoF was performed on each field to determine the critical linking length ($l_c$) that maximises the number of superclusters. A weighting function was then introduced to properly account for the selection function of clusters, i.e. that they are not homogeneously distributed in redshift.
We note that the use of a `tunable' linking length allows us to detect supercluster candidates at z$\geq$0.6, where the completeness of the sample drops, by assuming an additional density in order to maintain the value of the mean density similar to that of nearby clusters.
A 3D Voronoi tessellation technique \citep{Icke1987,Sochting2012} was then applied to clusters in both XXL fields in order to assess the reliability of the supercluster detection procedure described in the previous paragraph.
Finally, the results of this procedure were compared to the set of superclusters found in XXL Paper II with a different method and to the supercluster found among the XXL-100 brightest cluster sample in order to find whether the same structures are identified.

The focus of this work is XLSSsC N01, the largest supercluster identified in XXL Paper XX, with an extension of $\sim$ 2 deg in right ascension and $\sim$ 3 deg in declination; the coordinates of the centroid of the structure are RA=36.954$^\circ$, Dec=-4.778$^\circ$ and redshift $z$=0.2956. The redshift of the supercluster was spectroscopically confirmed  for the first time with MOS optical spectroscopy obtained with the 4.2m William Herschel Telescope \citep[][XXL Paper XII]{Koulouridis2016}. The supercluster is composed of 14 spectroscopically confirmed clusters, whose main properties are  described in Table \ref{table_XLSSsC-N01}: 9 are classified as C1, 3 are C2, while only 2 are classified as C3. X-ray temperatures, luminosities, virial radii, and masses are derived from scaling relations starting from X-ray count rates following the procedure presented in XXL Papers XX and XXII. The number of galaxies in each structure is obtained in Sect.~3. 
In the following, we  consider only C1 and C2 (C1+C2) clusters for which X-ray count-rates provide good-quality measurements of virial properties. Among these, we  exclude XLSSC 028 because it is located outside the region covered by our photo-z catalogue of galaxies (see Sec.~\ref{galaxy_catalog}).

We note that the XLSSsC N01 cluster $\rm M_{500,scal}$ masses range from $4 \times 10^{13}$ to over $2 \times 10^{14} M_\odot$, with half of the clusters having masses greater than $\rm 10^{14} M_\odot$, which corresponds to X-ray luminosities $L_{500,scal}^{XXL}$ greater than $10^{43}$ erg/s. The distribution of virial masses and X-ray luminosities does not differ from that of the overall C1+C2  sample analysed in XXL Paper XXII, meaning that, at first sight, clusters assembling  to form a supercluster do not have unusual virial masses or X-ray luminosities.

\begin{table*}
\caption{X-ray and membership properties of clusters within the XLSSsC N01 superstructure. Column 1 gives the IAU official name of clusters; column 2 is the classification of clusters according to the level of contamination as explained in XXL Paper II; column 3 is the spectroscopic redshift of the clusters; columns 4 and 5 contain the RA-Dec coordinates of the X-ray centres of clusters; columns 5--9 report all X-ray parameters derived through scaling relations from X-ray count-rates (XXL Paper XX): temperature (T$_{300kpc,scal}$), virial radius (r$_{500,scal}$), virial mass ($\rm M_{500,scal}$), luminosity ($\rm L_{500,scal}^{XXL}$). Velocity dispersion ($\sigma_{200}$) was measured in XXL Paper XXII using a relation based on the virial theorem given in \citealt{Poggianti2006}; columns 10 and 11 report the number of spectroscopic members within 1$\rm r_{200}$ and 3$\rm r_{200}$ as assigned in section \ref{environment}.\label{table_XLSSsC-N01}}
{\begin{tabular}{l}
\resizebox{\textwidth}{!}{\begin{tabular}{llllllllllll}
\hline
\multicolumn{1}{c}{XLSSC} &
\multicolumn{1}{c}{class} &
\multicolumn{1}{c}{$z$} &
\multicolumn{1}{c}{RA} &
\multicolumn{1}{c}{Dec} &
\multicolumn{1}{c}{T$_{300kpc,scal}$} &
\multicolumn{1}{c}{r$_{500,scal}$} &
\multicolumn{1}{c}{M$_{500,scal}$} &
\multicolumn{1}{c}{$L^{XXL}_{500,scal}$} &
\multicolumn{1}{c}{$\sigma_{200}$} &
\multicolumn{1}{c}{N$_{gal,1r200}$} &
\multicolumn{1}{c}{N$_{gal1-3r200}$} \\
 & & &
\multicolumn{1}{c}{(deg)} &
\multicolumn{1}{c}{(deg)} &
\multicolumn{1}{c}{(keV)} &
\multicolumn{1}{c}{(kpc)} &
\multicolumn{1}{c}{($10^{13} M_{\odot}$)} &
\multicolumn{1}{c}{$10^{42}$(erg/s)}&
\multicolumn{1}{c}{(km/s)} & & \\
\hline
008 & C1 & 0.2989 & 36.336 & -3.801 & 1.6$\pm$0.2 & 579$\pm$53 & 7$\pm$2& 5.5$\pm$0.9 & 404$^{+32}_{-38}$ & 12 & 6\\
013 & C1 & 0.3075 & 36.858 & -4.538 & 2.0$\pm$0.2 & 635$\pm$57 & 10$\pm$3 & 8.7$\pm$0.8 & 445$^{+37}_{-46}$ & 31 & 22\\
022 & C1 & 0.2932 & 36.917 & -4.858 & 3.1$\pm$0.2 & 835$\pm$79 & 22$\pm$6 & 30.1$\pm$1.2 & 588$^{+44}_{-53}$ & 38 & 50\\
024$^{\star}$ & C3 & 0.2911 & 35.744 & -4.121 & - & - & - & - & - & - & -\\
027 & C1 & 0.2954 & 37.012 & -4.851 & 2.4$\pm$0.2 & 710$\pm$64 & 13$\pm$4 & 14.1$\pm$1.1 & 494$^{+41}_{-51}$ & 15 & 13\\
028$^{\star}$ & C1 & 0.2969 & 35.984 & -3.098 & 1.5$\pm$ 0.2 & 545$\pm$52 & 6$\pm$2 & 4.1$\pm$0.9 & 380$^{+34}_{-43}$ & - & -\\
070$^{\star}$ & C3 & 0.3008 & 36.863 & -4.903 & - & - & - & - & - & - & -\\
088 & C1 & 0.2951 & 37.611 & -4.581 & 2.5$\pm$0.2 & 725$\pm$66 & 14$\pm$4 & 15.6$\pm$1.4 & 505$^{+40}_{-48}$ & 16 & 10\\
104 & C1 & 0.2936 & 37.324 & -5.895 & 2.5$\pm$0.2 & 735$\pm$67 & 15$\pm$4 & 16.5$\pm$1.4 & 512$^{+38}_{-45}$ & 7 & 35\\
140 & C2 & 0.2937 & 36.303 & -5.524 & 1.2$\pm$0.2 & 491$\pm$53 & 4$\pm$1 & 2.5$\pm$0.8 & 337$^{+23}_{-27}$ & 2 & 11\\
148 & C2 & 0.2938 & 37.719 & -4.859 & 1.8$\pm$0.2 & 608$\pm$63 & 8$\pm$3 & 6.8$\pm$1.8 & 423$^{+42}_{-55}$ & 11 & 23\\
149 & C2 & 0.2918 & 37.634 & -4.989 & 2.0$\pm$0.2 & 655$\pm$60 & 10$\pm$3 & 9.5$\pm$1.3 & 455$^{+37}_{-45}$ & 8 & 11\\
150 & C1 & 0.2918 & 37.661 & -4.992 & 2.2$\pm$0.2 & 678$\pm$62 & 12$\pm$3 & 11.2$\pm$1.3 & 472$^{+33}_{-39}$ & 9 & 0\\
168 & C1 & 0.2948 & 37.387 & -5.880 & 2.8$\pm$0.2 & 790$\pm$74 & 18$\pm$5 & 23.2$\pm$1.7 & 550$^{+42}_{-51}$ & 12 & 3\\
\hline
\end{tabular}}\\
$^\star${\scriptsize These clusters are excluded from this analysis either because they are classified as C3 or because of the lack of photometric data, as explained in the main text.}\\
\end{tabular}}
\end{table*}

\subsection{Galaxy catalogue} 
\label{galaxy_catalog}

To characterise the properties of the galaxies in the XLSSsC N01 supercluster, we extract the useful information from the spectrophotometric catalogue  presented by XXL Paper XXII. We focus on the area covered by the supercluster (RA~[35.25:38.0], Dec~[-6.25:-3.5]), and redshift range $0.25<z<0.35$.


The photometric and photo-z information for this region are mainly drawn from the CFHTLS-T0007 photo-z catalogue in the W1 Field ($\rm 8^\circ \times 9^\circ$, centred at RA=34.5$^\circ$ and Dec=-07$^\circ$). The data cover the wavelength range 3500\AA $< \lambda < 9400$\AA $\,$ in the $u^*$, $g^\prime$, $r^\prime$, $i^\prime$, and $z^\prime$ filters. These data are complemented with photometric and photo-z measurements in the SF catalogue (Sotiria Fotopoulou, private communication), containing aperture magnitudes in the $g^\prime$, $r^\prime$, $i^\prime$, $z^\prime$, $J^\prime$, $H^\prime$, and $K^\prime$ bands, which have been converted into total magnitudes using a common subsample of galaxies with the CFHTLS-T0007 W1 field catalogue.
The percentage of galaxies belonging to this sample that will be included in the scientific analysis presented in this paper is 2.5\%.

All magnitudes are Sextractor \verb!MAG_AUTO! magnitudes \citep{BertineArnouts1996} in the AB system corrected for Milky Way extinction according to \cite{Schlegel1998}.
The error associated with photo-z in the magnitude range that we are probing in this work has a dependence with redshift of $\sigma/(1+z) \sim 0.03$, and therefore assumes the minimum value of 0.039 at $z$=0.25 and the maximum 0.42 at $z$=0.35.


Spectroscopic redshifts are drawn from the XXL spectroscopic database hosted in the CeSAM (Centre de donn\'eeS Astrophysiques de Marseille) database in Marseille.\footnote{http://www.lam.fr/cesam/} 
As described in XXL Paper XXII, the starting point is a heterogeneous ensemble of spectra and redshift coming from different surveys superposed in the sky (mainly GAMA, SDSS, VIPERS, VVDS, VUDS, and XXL dedicated spectroscopic campaigns, see Table 2 in XXL Paper XXII), and the final spectroscopic catalogue was obtained by removing duplicates from the database using a careful combination of selection criteria (called the  priorities) that take into account the  parent survey of the spectrum and the quality of the redshift measurement.
Overall, the uncertainties on the galaxy redshift in the database vary from 0.00025 to 0.0005, computed from multiple observations of the same object and depending on the sample used (more details on the XXL spectroscopic database are given in XXL Paper XX); we  consider the highest value in this range as the typical redshift error for all objects.

A further 17 spectroscopic redshifts (Lonoce, private communication), obtained observing the centre of the XLSSsC N01 field with the AF2 multifiber spectrograph at the 4.2m William Herschel Telescope (WHT, La Palma
Island, Spain) in January 2017 during a campaign within the WEAVE project, appear to confirm the complexity of the supercluster. These data will be presented in Lonoce et al. (in prep.).\footnote{see http://www.ing.iac.es/weave/science.html for more information on the WEAVE project.}

The  spectroscopic catalogue of galaxies in the area of the supercluster and in the redshift range $0.25 \leq z \leq 0.35$ contains 4057 galaxies. The resulting spectrophotometric catalogue (with matching spectroscopic and photometric information) contains 3759 objects. 

\subsection{Tools}

To derive the properties of galaxies and of their stellar populations, we exploit two different codes.


Absolute magnitudes are computed using LePhare\footnote{http://www.cfht.hawaii.edu/~arnouts/lephare.html} \citep{Arnouts1999,Ilbert2006}, as described in XXL Paper XXII. This code was developed mainly to compute photometric redshifts, but it can also compute physical properties of galaxies, and the spectroscopic redshift can be used as an input fixed parameter to improve the quality of the physical outputs.
The LePhare output physical parameters that are going to be used in this work are absolute magnitudes, and thus rest-frame colours.

Galaxy stellar population properties have been derived by fitting the spectra with \textsc{SINOPSIS}\footnote{http://www.crya.unam.mx/gente/j.fritz/JFhp/SINOPSIS.html} (SImulatiNg OPtical Spectra wIth Stellar population models), a spectrophotometric model fully described in \citet{Fritz2007, Fritz2011,Fritz2017} and already largely used to derive physical properties of galaxies in many samples  \citep{Dressler2009,Vulcani2015,Guglielmo2015,Paccagnella2016,Paccagnella2017,Poggianti2017}. 
It is based on a stellar population synthesis technique that reproduces the observed optical galaxy spectra. 
All the main spectrophotometric features are reproduced by summing the theoretical spectra of simple stellar populations of 12 different ages (from 3 $\times 10^6$ to approximately 14$\times 10^9$ years).

Among other properties, the code provides estimates of star formation rates (SFRs), stellar masses ($M_*$) and stellar ages (the luminosity-weighted and mass-weighted age). 
For a given single stellar population (SSP), we recall that three different kinds of stellar mass can be distinguished \citep{Renzini2006,Longhetti2009}: M1 is the initial mass of the SSP at age zero,  which is nothing but the mass of gas turned into stars; M2 is the mass locked into stars, both those which are still in the nuclear-burning phase, and remnants such as white dwarfs, neutron stars, and stellar black holes; and M3 is the mass of stars in the nuclear-burning phase.
Hereafter, we  use stellar mass values corresponding to the M2 definition.

We run \textsc{SINOPSIS} on the subsample of spectra provided by SDSS and GAMA, which are flux calibrated and have the best available spectral quality, and which are the main contributors to the final sample that will be used in this paper (see Section \ref{completeness_subsect}).

\subsection{Spectroscopic completeness}
\label{completeness_subsect}
Each galaxy in the sample is weighted for spectroscopic incompleteness as computed in XXL Paper XXII.
Briefly, the whole XXL-N area (where the supercluster is located) is divided into several cells made of three stripes in declination, and  right ascension intervals of 1 deg width. The completeness ratio is computed as the number of galaxies in the spectrophotometric catalogue divided by the number of galaxies in the photometric catalogue in magnitude bins with an amplitude of 0.5 mag in the observed $r^\prime$ band. Further details about the completeness correction can be found in XXL Paper XXII.
The spectroscopic completeness analysis described in XXL Paper XXII was conducted on the spectrophotometric sample with LePhare absolute magnitude estimates. Given that in this paper we also make use  of the subsample with the outputs from \textsc{SINOPSIS}, we verified that the spectroscopic completeness relative to this sample did not vary in the whole XLSSsC N01 region, and also that the completeness curves of the galaxies within clusters are statistically similar to those of galaxies in the field.

\begin{figure*}
\begin{center}
\includegraphics[scale=0.5]{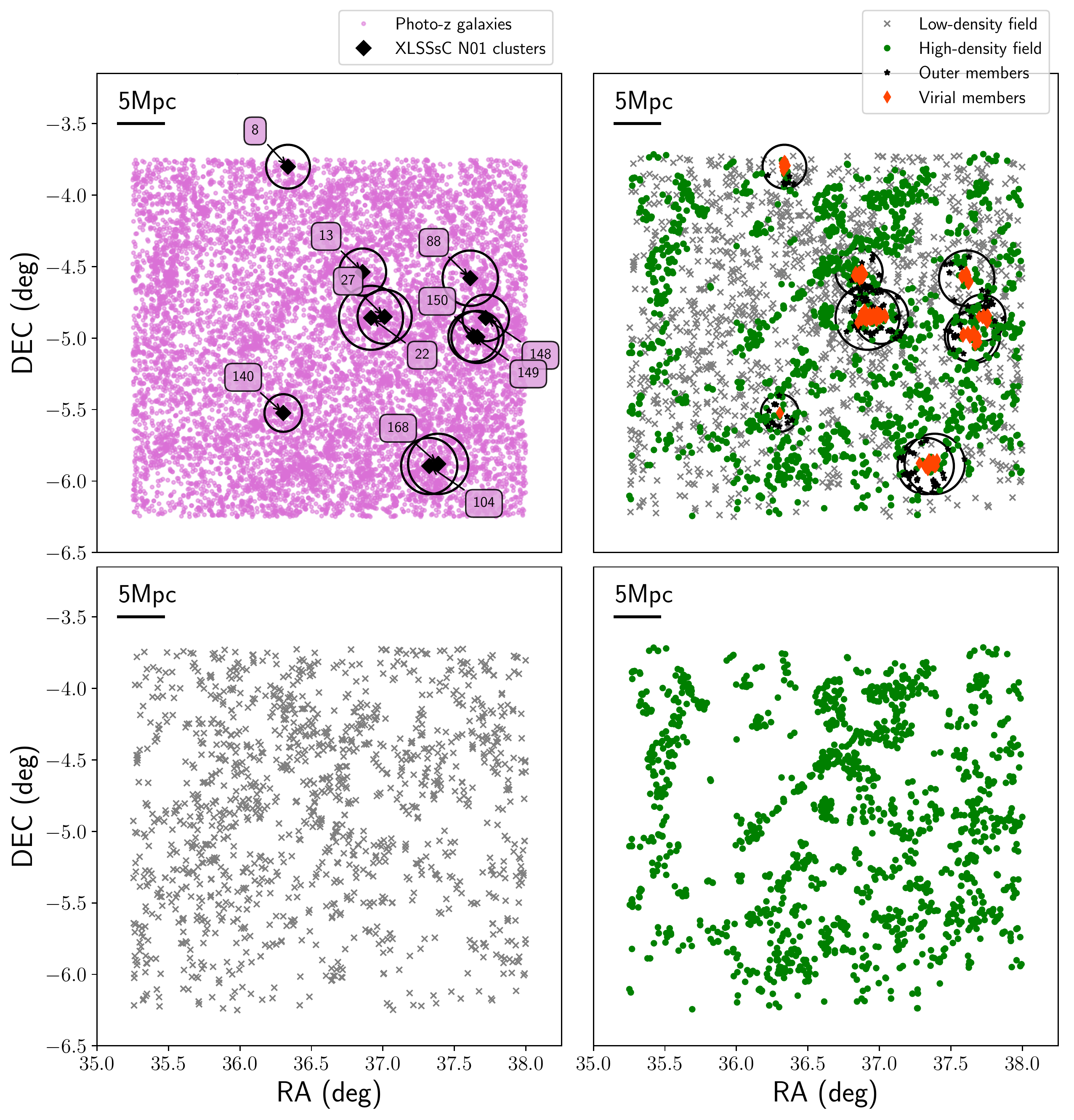}
\caption{Sky distribution of galaxies in the XLSSsC N01 supercluster region. Top left:  Galaxies with a photo-z redshift in the range between 0.25 and 0.35 and used to compute the LD (black points). Top right: Galaxies with a spectroscopic redshift, colour-coded according to their environment (see Sect. \ref{environment}). Grey crosses are low-density field galaxies, green dots are high-density field galaxies, dark orange diamonds are virial members, and black stars are outer members. In  the top panels, black circles show the projected extension in the sky of 3 $\rm r_{200}$ for each cluster in the superstructure. The two bottom panels show the low- and high-density field samples separately, with the same symbols as the top right panel.}
\label{RA_DEC_E01_members_flagenv_photoz}
\end{center}
\end{figure*}

We apply the same magnitude cut and completeness weight to all samples and environments. Following XXL Paper XXII, the magnitude completeness limit is set to $r=20.0$; at the redshift of XLSSsC N01, this corresponds to an absolute magnitude of $M_r\sim -21.4$.
In the magnitude complete sample, which selects 2429 galaxies out of 3759, 97\% of the galaxies come from the GAMA and SDSS surveys (2323 and 27 galaxies, respectively), which provide the spectra analysed in the following sections.
The magnitude completeness limit is converted into a mass completeness limit following the procedure detailed in XXL Paper XXII and summarised  below.
The stellar mass completeness limit computation is performed in the magnitude complete sample ($r\leq 20.0$); it  is strongly redshift dependent, so to compute it we divided our entire redshift range into several intervals. In each redshift interval, we built $(g-r)_{rest-frame}$ versus $r$ rest-frame  colour-magnitude diagram (CMD) and computed the stellar mass completeness limit following \cite{Zibetti2009},
\begin{footnotesize}
\begin{equation}
\mathcal{M}_{lim,M_{\odot}}=-0.840+1.654(g-r)_{rest-frame,lim} + 0.4(M_{r,\odot} - M_r)
,\end{equation}
\end{footnotesize}
where $(g-r)_{rest-frame,lim}$ is rest-frame colour limit defined as the colour of the reddest galaxy in the CMD excluding outliers, M$_{r,lim}$ is the absolute r-band magnitude of the faintest galaxy in the region of the CMD close to the rest-frame colour limit, and the absolute magnitude of the Sun is $\rm M_{r,\odot} = 4.64$.
At the redshift of the  XLSSsC N01 supercluster and in the entire redshift range adopted in this paper, we assume a conservative stellar mass completeness limit of $\rm log(M_{\ast,lim}/M_{\odot}) = 10.8$.

\section{Characterisation of the XLSSsC N01 supercluster: the definition of environment}

\label{environment}

\begin{figure}
\includegraphics[scale=0.45]{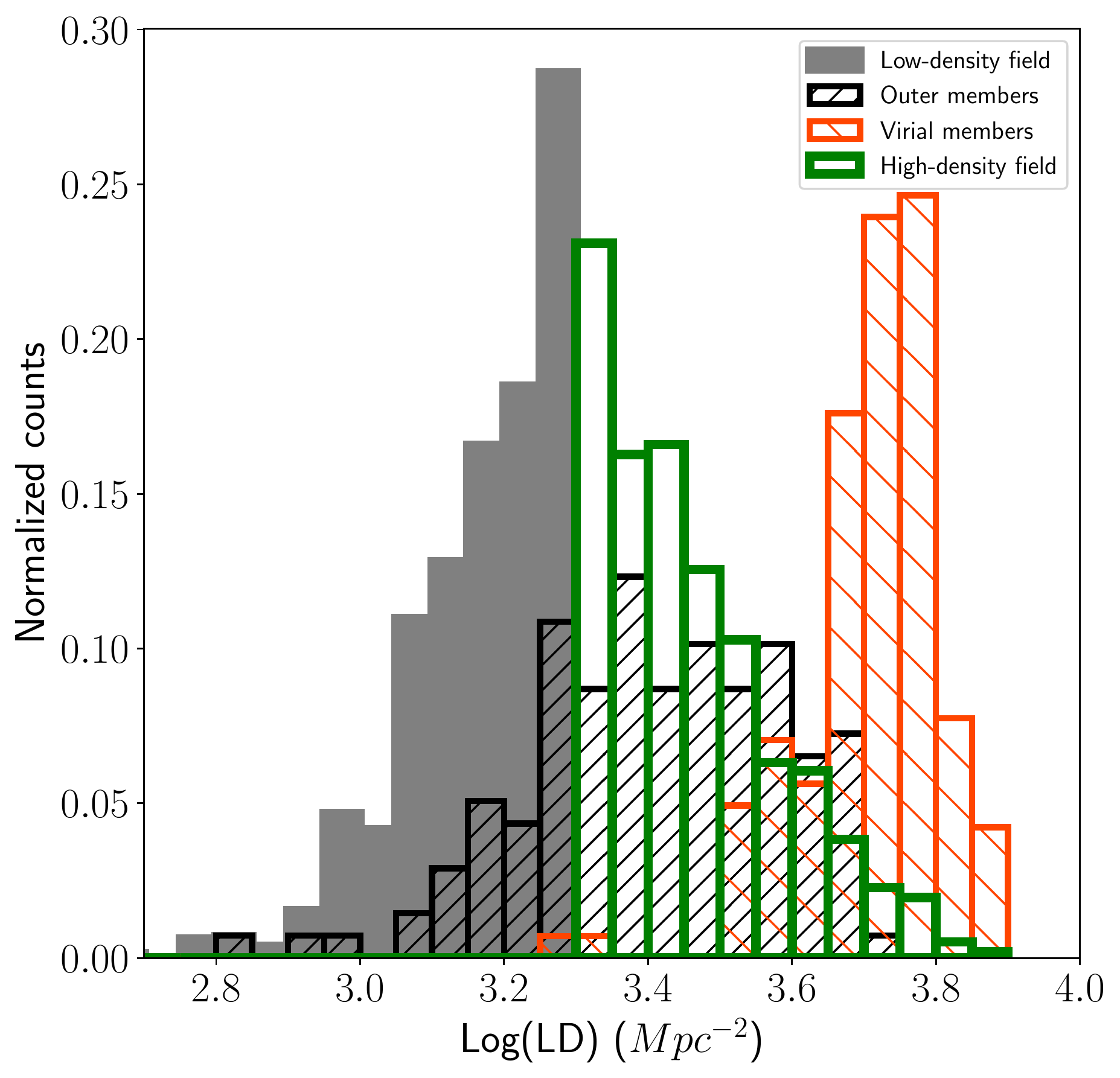}
\caption{Normalised local density distribution  computed using the photo-z sample in redshift range $0.25 \leq z_{phot} \leq 0.35$.  The grey histogram represents the low-density field, the green empty histogram  galaxies in the high-density field, the black hatched histogram the cluster outer members, and the orange hatched histogram the cluster virial members.}
\label{histo_ld}
\end{figure}

\begin{figure*}
\begin{center}
\includegraphics[scale=0.45]{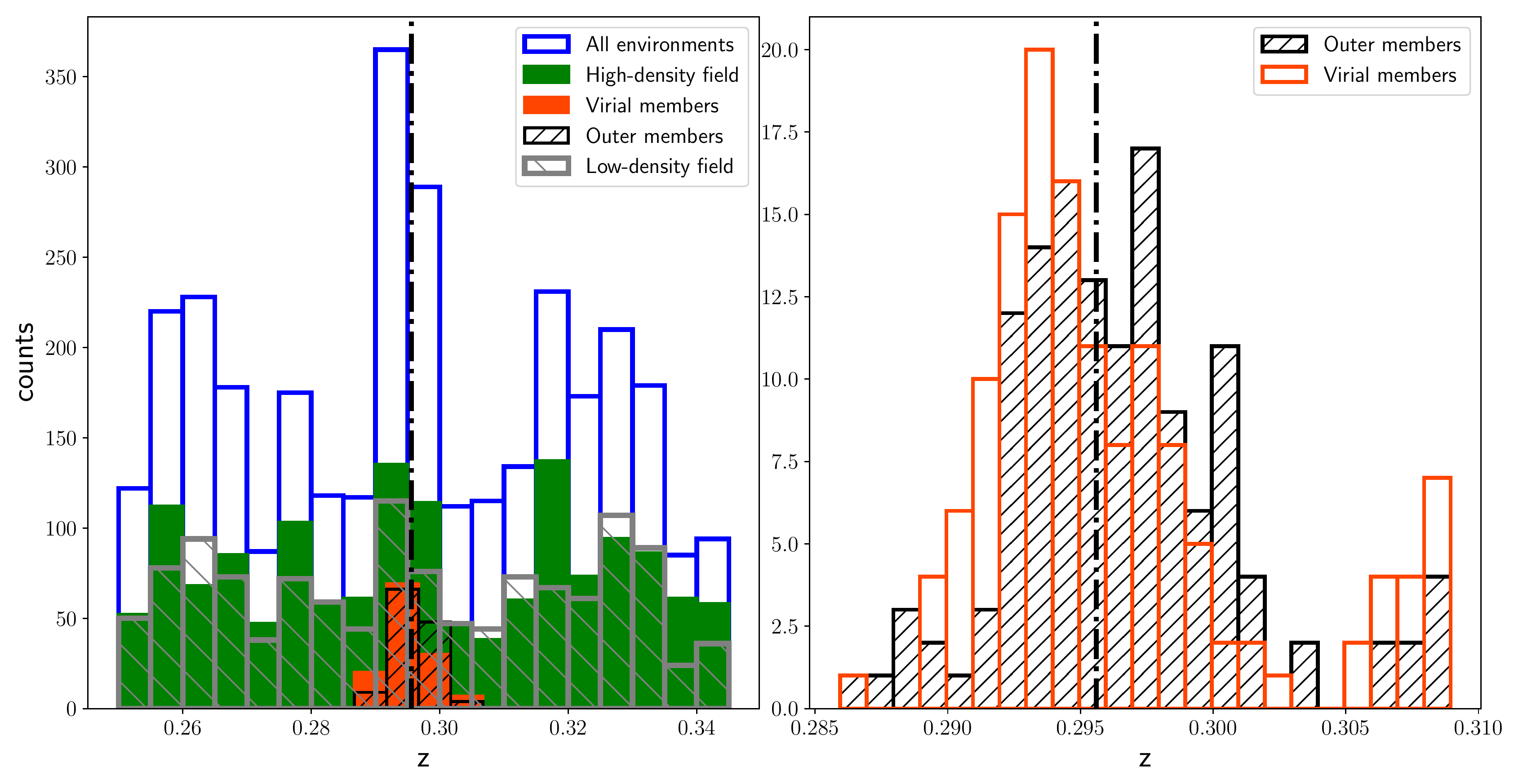}
\caption{Redshift distribution of the spectrophotometric sample in the region including the  XLSSsC N01 supercluster. The centroid redshift of the supercluster is represented with a black dashed line in both panels. Left panel: Whole spectroscopic sample (blue histogram),  low-density field galaxies (grey distribution),  galaxies in the high-density field (green distribution), and  galaxies classified as virial and outer members (dark orange and black histogram, respectively). Right panel: Zoom in on the  virial and outer members, with the same colours used in the left panel (see  Sect. \ref{environment} for details about the definition
of different environments).}
\label{z_E01_allenv_members}
\end{center}
\end{figure*}

In this section, we consider Fig. \ref{RA_DEC_E01_members_flagenv_photoz} as a reference for the spatial distributions of XLSSsC N01 clusters and galaxies. The upper panels of the figure show the centre of each cluster belonging to XLSSsC N01, along with $3\times \rm r_{200}$ circles indicating the cluster spheres of influences.
The spectrophotometric sample of galaxies in the top right panel shows, with different colours, galaxies belonging to the four environments we  define in this section. We note that at this stage we are considering the spectrophotometric sample with LePhare outputs presented in the previous sections with a cut at observed magnitudes fainter than the spectroscopic completeness limit, i.e. r=22.0, which is the most suitable  and contains 3120 galaxies; thus, the numbers and fractions written in this section refer to this sample, unless otherwise stated.
Based on virial properties of clusters together with redshifts of galaxies and their distance from the clusters within XLSSsC N01, we distinguish two membership regions: 
\begin{itemize}
\item Cluster virial members are galaxies whose spectroscopic redshift lies within $\rm 3 \sigma$ of their cluster mean redshift, where $\rm \sigma$ is the velocity dispersion of their host cluster, and whose projected distance from the cluster centre is $< 1 \, r_{200}$.\footnote{We derive r$_{200}$ from r$_{500,scal}$ by simply dividing the latter by 0.7 according to the relation adopted in \citet{Ettori2008}.} The number of cluster virial members is 130 (4.2\% of the spectrophotometric sample). Virial members are marked in the top right panel of Figure \ref{RA_DEC_E01_members_flagenv_photoz} with dark orange diamonds.
\item Cluster outer members are galaxies whose spectroscopic redshift lies within $\rm 3 \sigma$ from their cluster mean redshift, and whose projected distance from the cluster centre is between 1 and 3 $r_{200}$.
The number of cluster outer members is 133 (4.3\% of the spectrophotometric sample), and they are marked in the top right panel of Figure \ref{RA_DEC_E01_members_flagenv_photoz} by black stars.
\end{itemize}

In order to characterise galaxies in different environments, we extend the definitions of cluster virial and outer members as follows.
We consider the redshift range $0.25 \leq z \leq 0.35$, and we remove galaxies belonging to other nine clusters in the same region, which are not members of the supercluster because of their redshift: 197 galaxies  of the 2857  galaxies (6.9\%) which are neither in the virial nor in the outer membership regions of XLSSsC N01 clusters are identified as members of these nine clusters and are removed from the sample, so that the remaining 2660 galaxies belong to a {\it general field} sample not contaminated by the presence of other X-ray clusters.

Then, we compute projected local densities (LD) in order to further refine the {\it general field environment}.
We consider all galaxies in the photo-z sample with an observed magnitude r $\leq$ 22.0, and a photo-z in the range $0.25 \leq z_{phot} \leq 0.35$: r=22.0 is the faintest magnitude at which the error in the photo-z estimate is sufficiently low ($\sigma/(1+z) \sim 0.03$, as reported in Sect. \ref{galaxy_catalog}), while the photo-z range is chosen on the basis of scatter in the spectroscopic versus photometric redshift plane in order to simultaneously minimise the contamination from galaxies with a photo-z within the selected range but with spectroscopic redshift outside of this range, and maximise the number of galaxies at the redshift of the supercluster. We include in the LD computation also galaxies with no reliable photo-z, but whose spectroscopic redshift is 0.25 $\leq$ spec-z $\leq$ 0.35.
The photo-z sample of galaxies used in the LD computation is shown in the top left panel of Figure \ref{RA_DEC_E01_members_flagenv_photoz}.
We define the projected LD relative to a given galaxy as the number of neighbours in a fixed circular region in the sky of radius 1 Mpc at z=0.2956 (the redshift of XLSSsC N01). We consider all galaxies in the photo-z sample in a slightly larger rectangular region with respect to that defined in Section \ref{environment} in order to minimise the regions in which boundary corrections had to be performed: $35.0 \leq RA (deg) \leq 38.25$, $-6.5 \leq Dec (deg) \leq -3.5$. 

The LD is defined as the ratio between the number counts of galaxies $N_c$ in the circle around the considered galaxy and the area $A$ of the circle itself.
Count corrections are performed for galaxies in the proximity of the edges in the high declination side of the rectangle (-3.76743 $\leq$ Dec (deg) $\leq$ -3.70524) by computing the area of the circular segment that falls outside the field and dividing the LD estimate by the ratio $F_c \, (\leq 1)$ between the area actually covered by the data and the circular area.

Figure \ref{histo_ld} shows the distribution of the projected LD of our sample. The distribution of cluster virial members is distinct and shifted towards higher values than that for the general field, while the distribution of outer members is broadened in the range of log(LD). 

For the general field, we separate galaxies whose log(LD) is higher and lower than the median value of the distribution ($\rm \log(LD/Mpc^{-2})\gtrsim$ 3.3). The former constitute the `high-density' field, the latter the `low-density' field sample. The number of galaxies belonging to the high-density field is 1436 (46.0\%) and to the low-density field is 1224 (39.2\%).

The low- and high-density field samples are given together in the top right panel of Figure \ref{RA_DEC_E01_members_flagenv_photoz} along with virial/outer members, and separately in the two bottom panels of the same figure to better visualise  their definitions.  The high-density field sample traces the presence of several structures around the clusters belonging to XLSSsC N01.

The redshift distributions of  galaxies  in the different environments is shown in the left panel of Figure \ref{z_E01_allenv_members}. 
The right panel of the same figure zooms in the redshift distribution of the members, highlighting how the distribution is bimodal, with a second peak in redshift that matches the XLSSC 013 cluster.

From Figure \ref{RA_DEC_E01_members_flagenv_photoz}, it can be seen that several groups and clusters are gathered into substructures within the supercluster region.

\begin{figure*}
\begin{center}
\begin{subfigure}{\textwidth}
\includegraphics[width=0.33\linewidth]{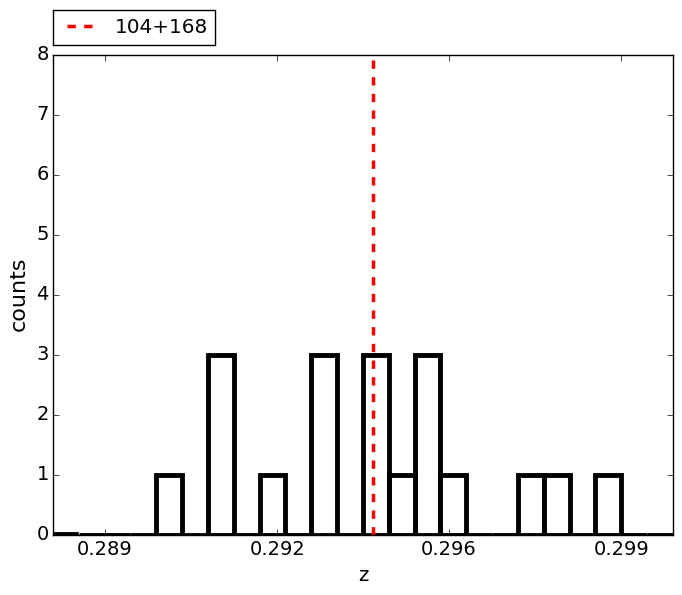}
\includegraphics[width=0.5\linewidth]{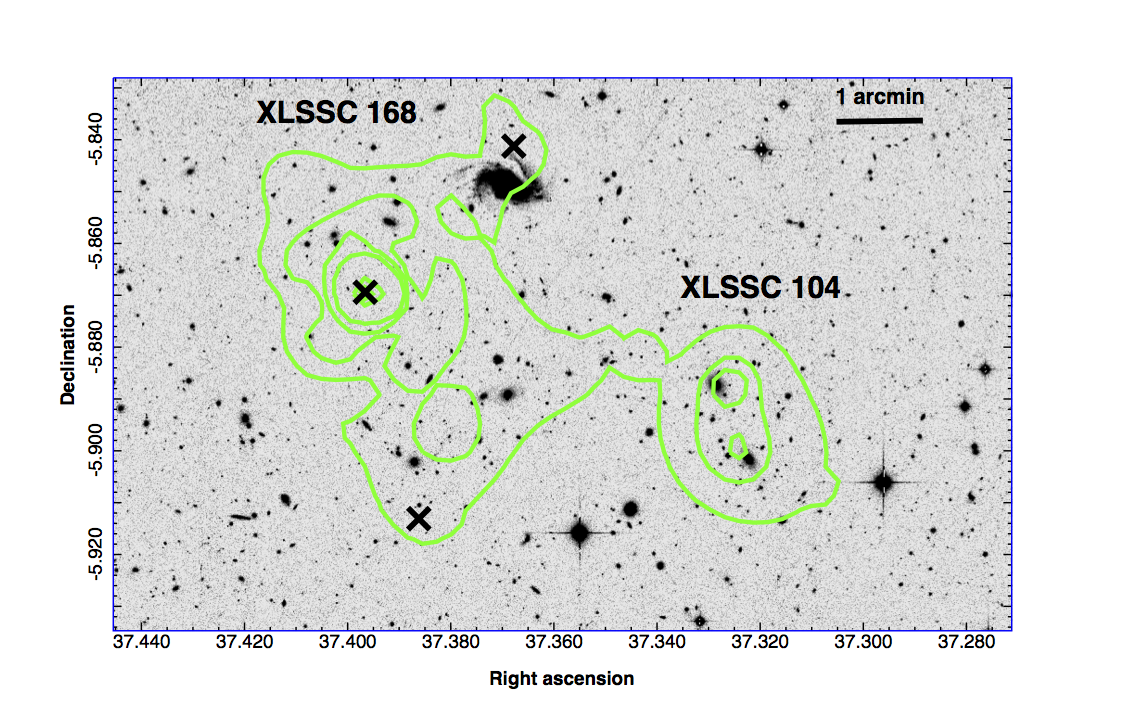}
\includegraphics[width=0.33\linewidth]{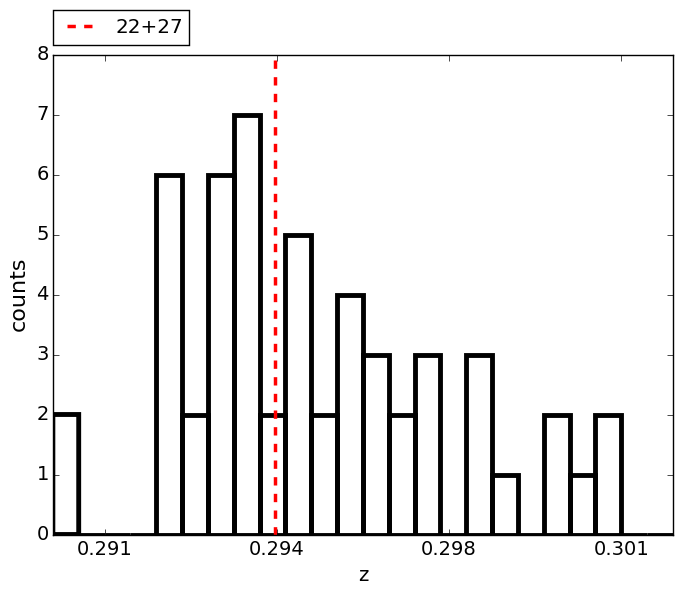}
\hspace{-2mm}
\includegraphics[width=0.35\linewidth]{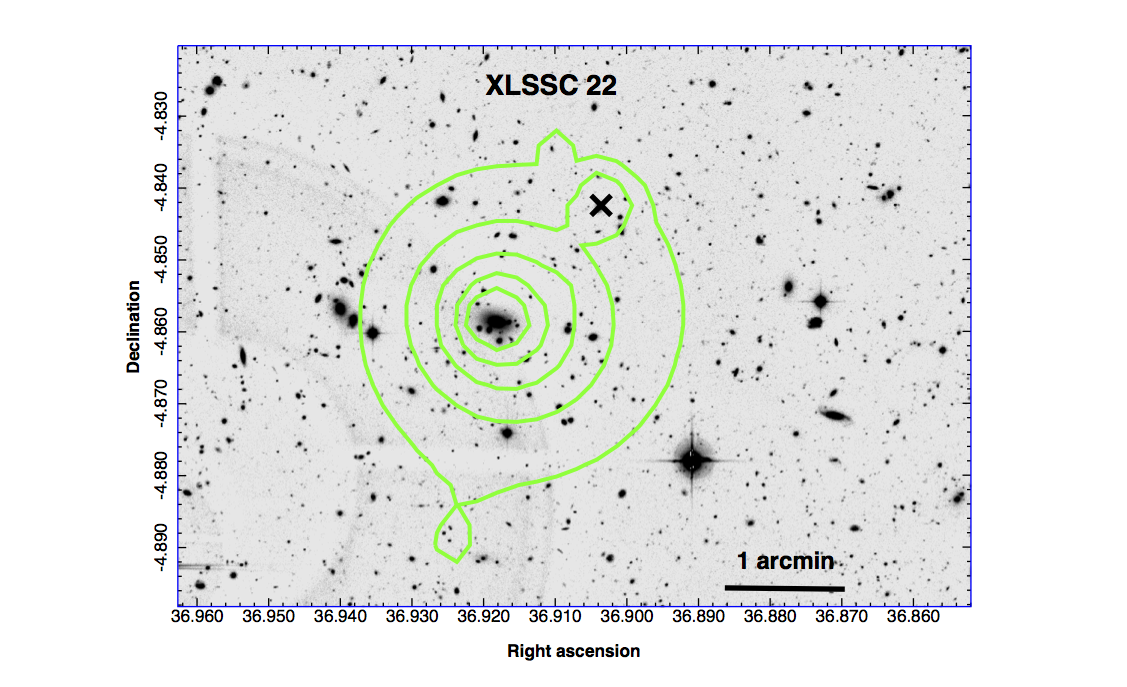}
\hspace{-10mm}
\includegraphics[width=0.35\linewidth]{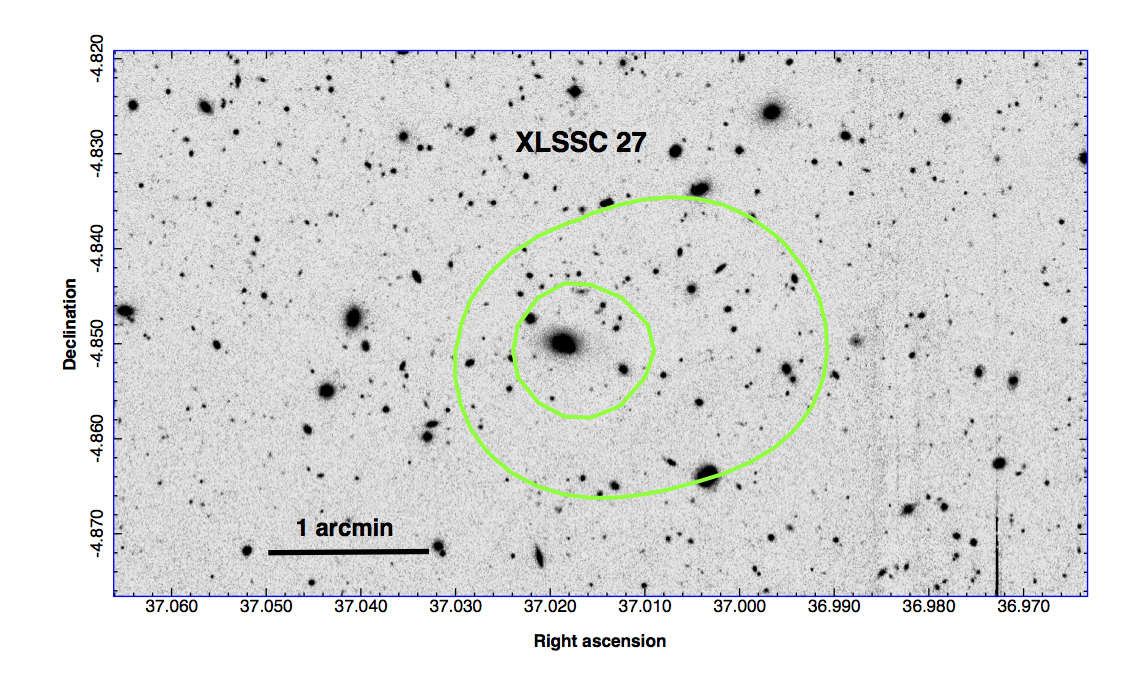}
\includegraphics[width=0.33\linewidth]{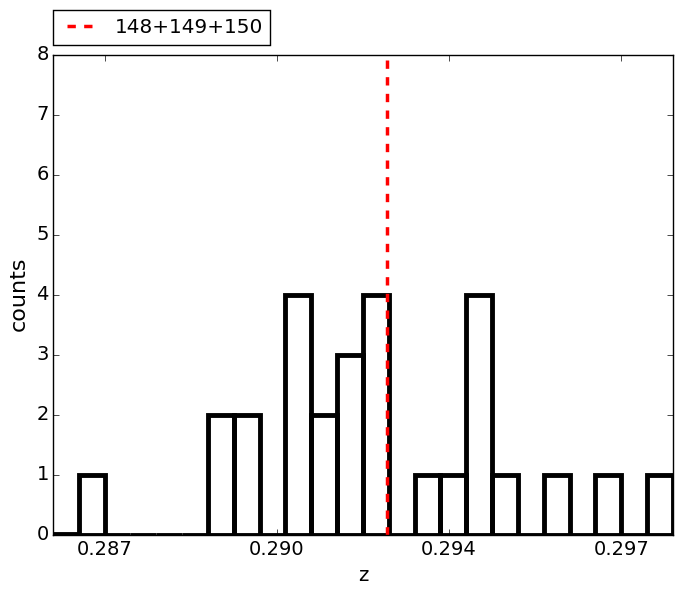}
\hspace{-2mm}
\includegraphics[width=0.35\linewidth]{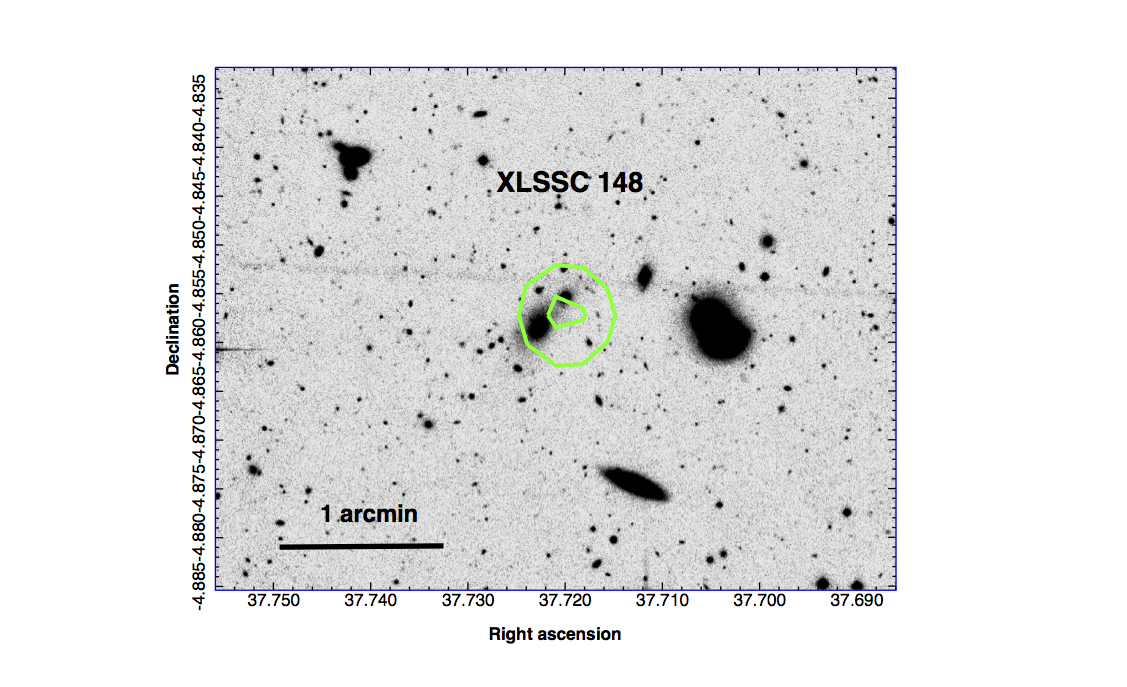}
\hspace{-10mm}
\includegraphics[width=0.35\linewidth]{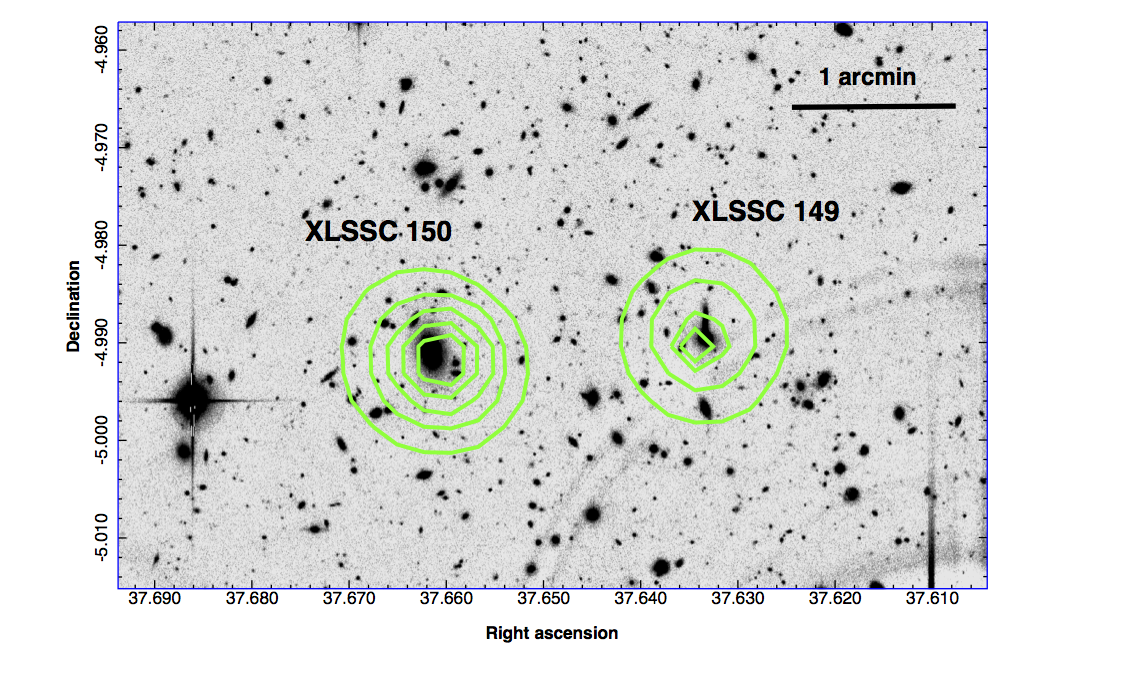}
\includegraphics[scale=0.33]{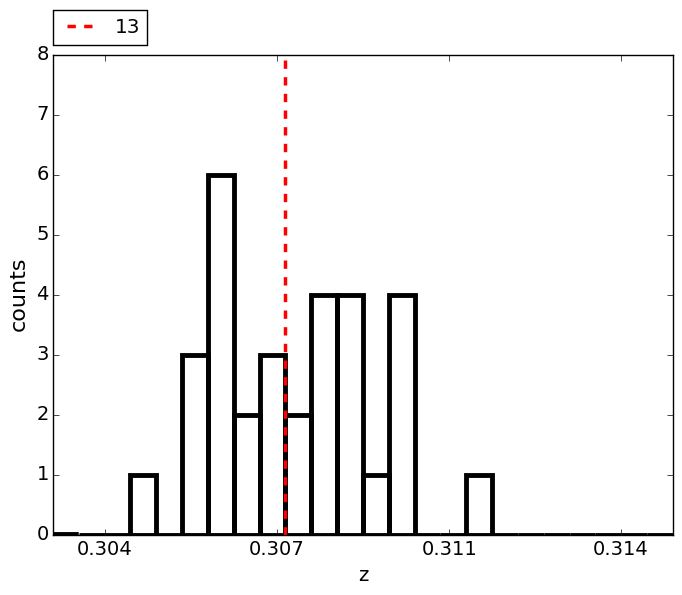}
\includegraphics[scale=0.45]{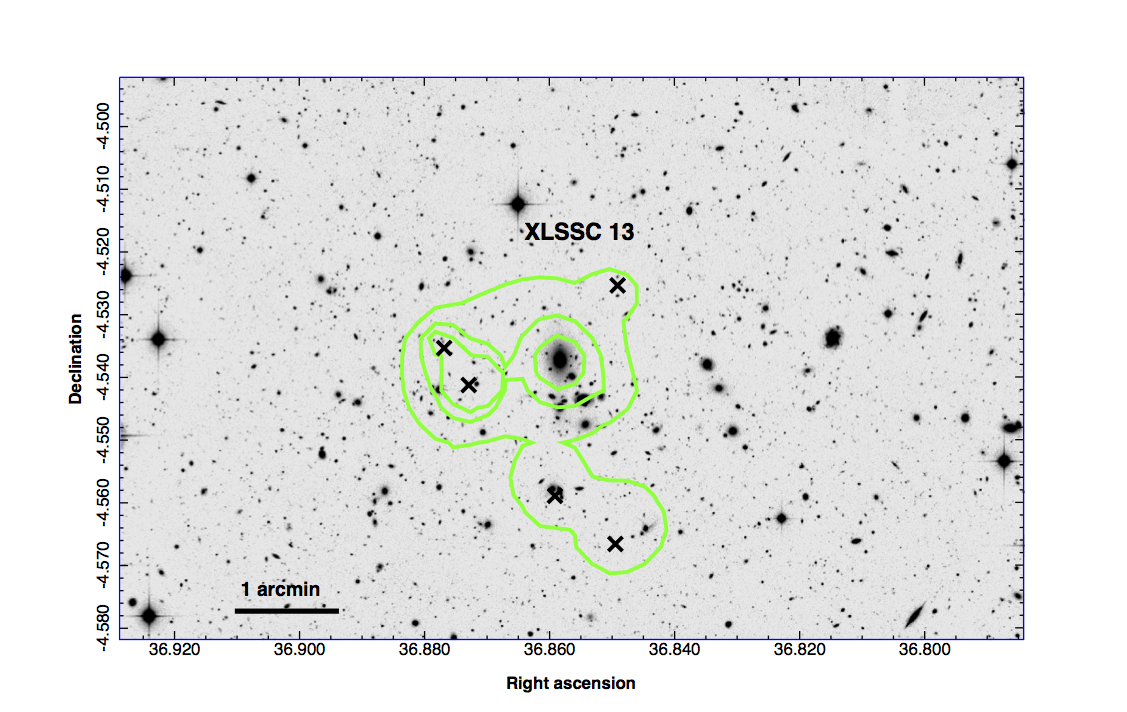}
\caption{Redshift distribution of XLSSsC N01 cluster virial members divided into seven substructures, and the relative X-ray contours. Left panels: Redshift distribution of the virial members in each substructure, as indicated in the labels. The redshift binning is the same in all histograms, and the x-axis extension depends on the redshift range covered by each substructure, for a better visualisation. 
The mean redshift of the substructure is shown with a vertical red dashed line. Right panels:  CFHTLS i-band image of the region surrounding the structures with X-ray contours superposed in green. Black crosses indicate the centre of the X-ray emission of point sources. The physical extension of the area in the sky is indicated within each panel.}
\label{xlssc_subpanels}
\end{subfigure}
\end{center}
\end{figure*}

\begin{figure*}\ContinuedFloat
\begin{center}
\begin{subfigure}{\textwidth}
\includegraphics[scale=0.33]{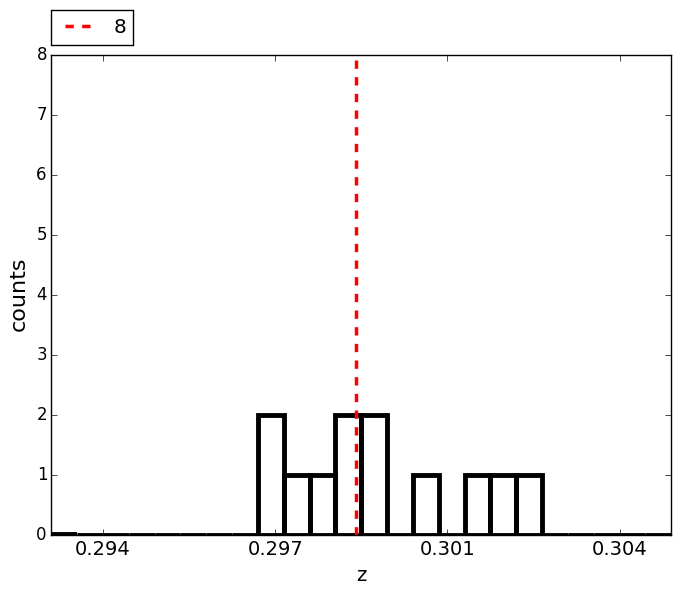}
\includegraphics[scale=0.45]{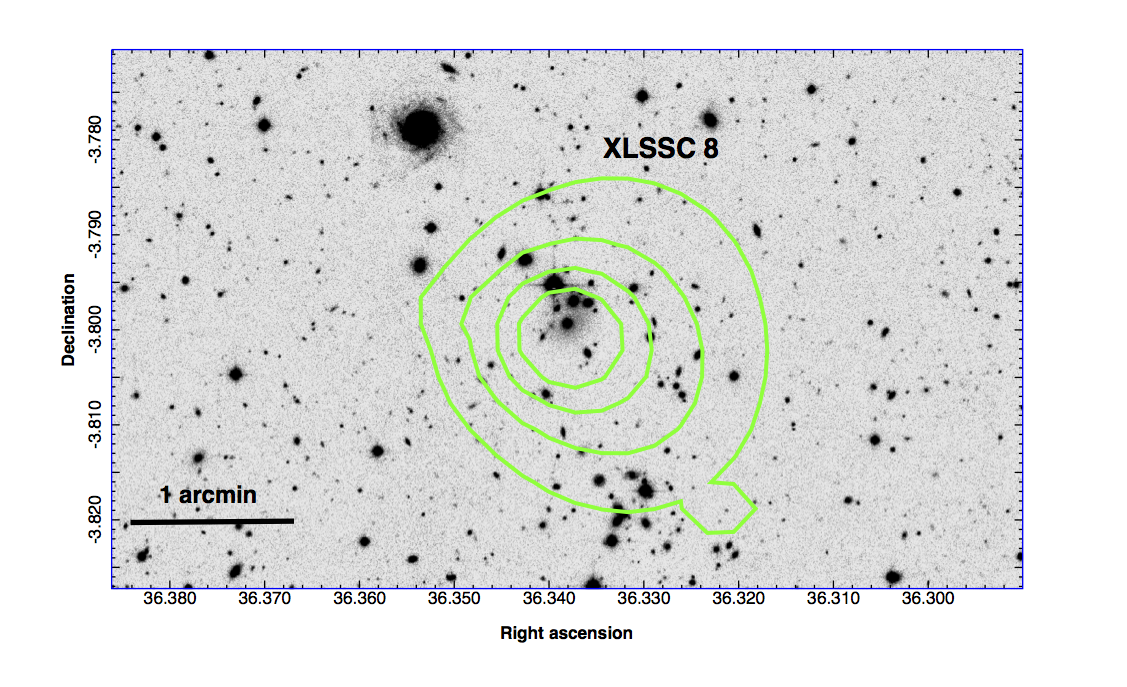}
\end{subfigure}
\begin{subfigure}{\textwidth}
\includegraphics[scale=0.33]{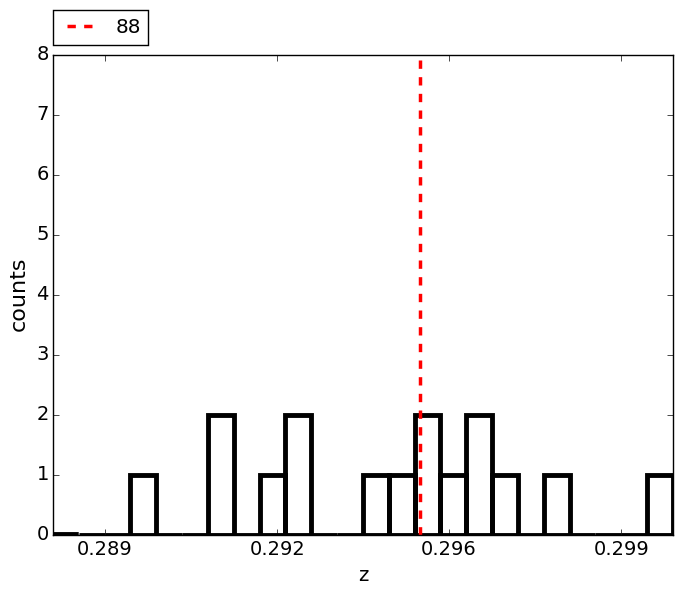}
\includegraphics[scale=0.45]{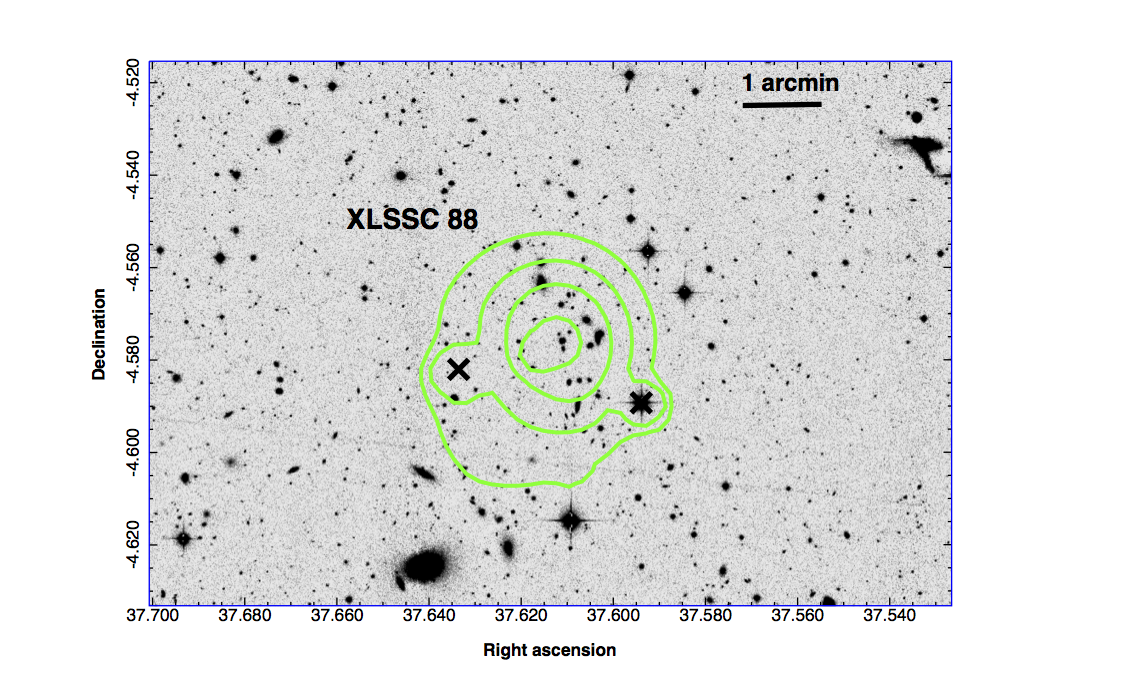}
\end{subfigure}
\begin{subfigure}{\textwidth}
\includegraphics[scale=0.33]{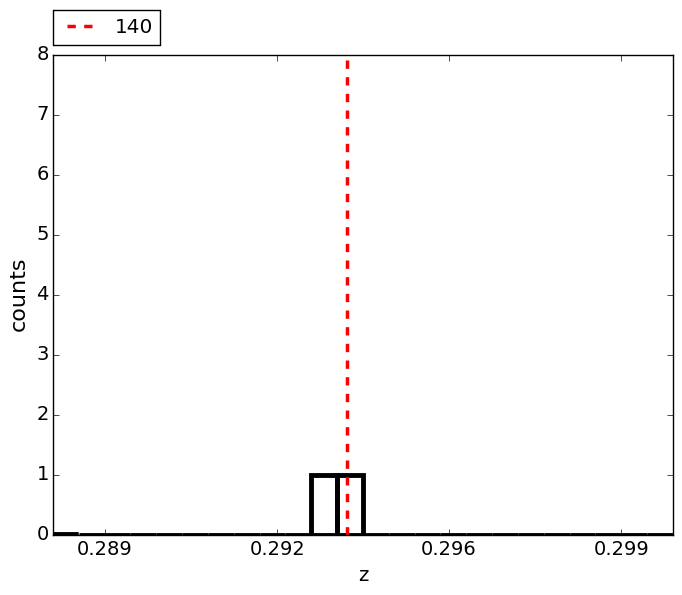}
\includegraphics[scale=0.45]{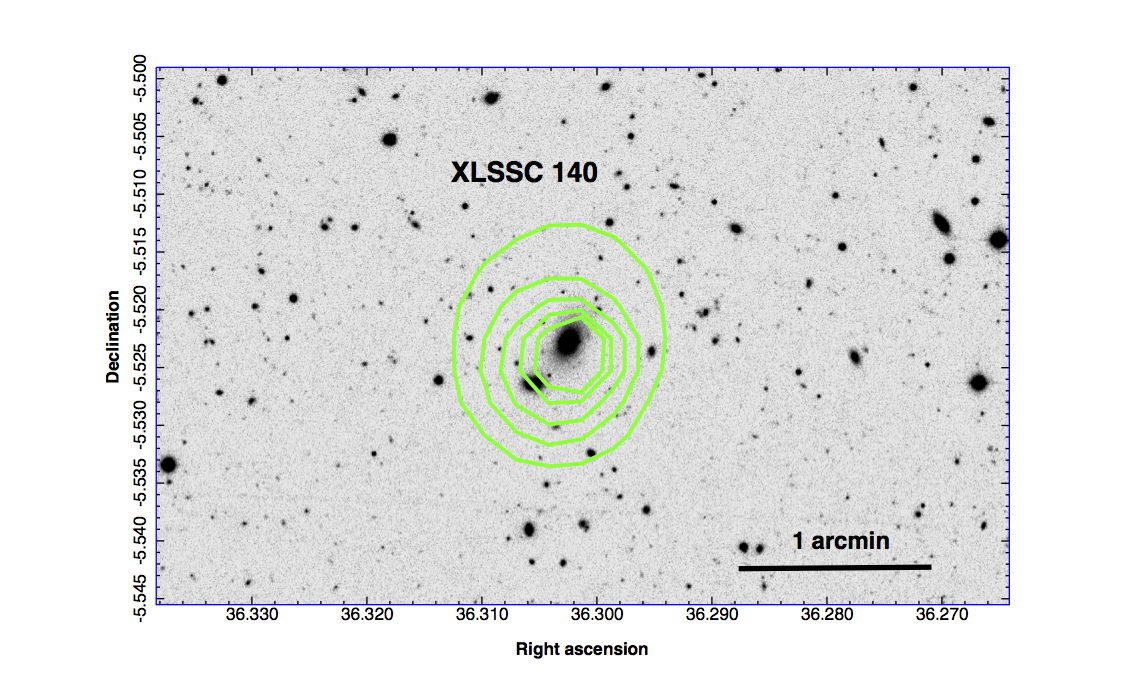}
\caption{Same as Figure \ref{xlssc_subpanels}, but for another set of substructures}
\end{subfigure}
\end{center}
\end{figure*}

The redshift distribution of cluster virial members is arbitrarily divided into seven substructures according to their position in the sky and is shown in Figure \ref{xlssc_subpanels}.
The motivation for this subdivision is twofold: first, groups and clusters within each substructure have overlapping virial radii and redshift distributions of member galaxies, which can be assigned to more than one cluster in many cases. In some cases, these clusters (e.g. XLSSC 104 and 168) also share X-ray contours. Second, the subdivision also aims to maximise the visibility of the X-ray contours related to each group and cluster (or substructure), as shown in the figure. For instance, XLSSC 008, 013, 088, and 140 are more distant from the other clusters that are shown together and therefore we show them separately.

\begin{table*}
\centering
\begin{tabular}{l|cc|cc}
\hline
\multicolumn{1}{c|}{} &
\multicolumn{2}{c|}{LePhare sample} &
\multicolumn{2}{c}{\textsc{ SINOPSIS} sample} \\
\multicolumn{1}{c|}{} &
\multicolumn{1}{c}{$r \leq 20$} &
\multicolumn{1}{c|}{$\rm \log(M_\ast/M_{\odot})\geq 10.8 $} &
\multicolumn{1}{c}{$r \leq 20$} &
\multicolumn{1}{c}{$\rm \log(M_\ast/M_{\odot})\geq 10.8 $} \\
\hline
Virial members & 75 (96) & 62 (76) & 70 (84) & 48 (59)\\
Outer members & 99 (120) & 61 (74) & 100 (120) & 59 (71)\\
High-density field & 1159 (1427) & 633 (746) & 1215 (1470) & 607 (724)\\
Low-density field & 958 (1189) & 450 (533) & 1024 (1252) & 438 (526) \\
\hline
\end{tabular}
\caption{Number of galaxies in the different environments above the magnitude and mass completeness limit, respectively, for the sample with successful fits from LePhare and \textsc{SINOPSIS}.  Numbers in parentheses are weighted for spectroscopic incompleteness.\label{N_env_table}}
\end{table*}

The final samples of galaxies that will be used in the scientific analysis is presented in Table~\ref{N_env_table}. We list the number of galaxies in the spectrophotometric catalogue with LePhare and \textsc{SINOPSIS} outputs, respectively, in all the environments defined in this section, both in the magnitude-limited and in the mass-limited samples.
We  make use of both catalogues because SFRs are available only for the \textsc{SINOPSIS} sample, while the LePhare sample
maximises the number of galaxies classified as cluster members.

\section{Stellar population properties versus environment}
\label{results}

\begin{figure}
\includegraphics[scale=0.375]{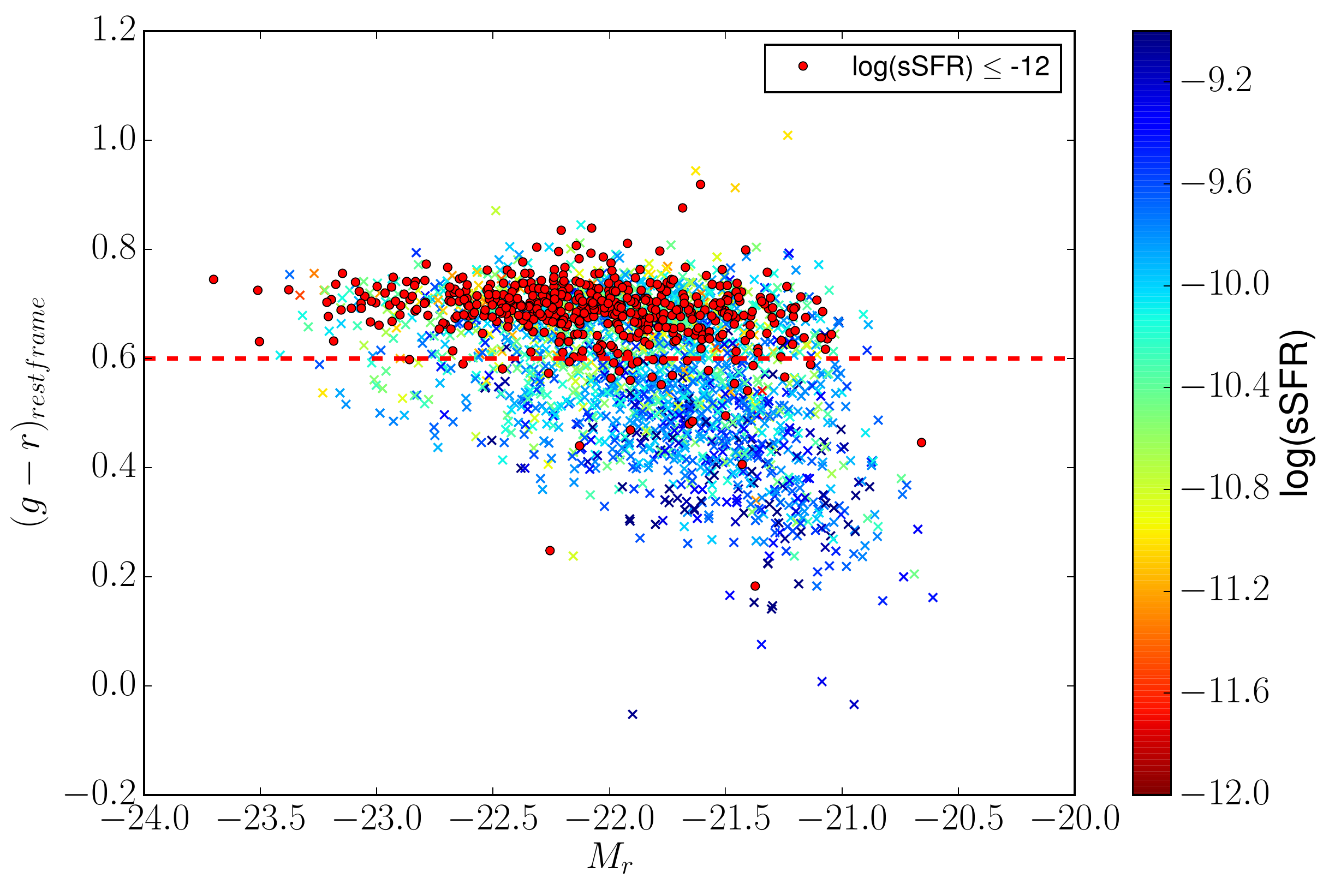}
\caption{Colour-magnitude diagram for galaxies in the magnitude-limited sample for the subset with both \textsc{SINOPSIS} and LePhare outputs. Red points indicate passive galaxies, while galaxies with log(sSFR)>-12 are colour-coded according to their sSFR. The red dotted line shows the separation between red and blue objects.}
\label{cmd_abs_g-r_LePh}
\end{figure}

In this section we present the analysis of the stellar population properties of galaxies in different environments using both the magnitude complete sample ($r \leq 20.0$) and the mass-limited sample ($\rm log(M_{\ast}/M_{\odot}) \geq 10.8$).

To distinguish between galaxies at different stages of their evolution, we exploit two different definitions of star-forming and passive galaxies. 
The first definition is based on the current SFR and stellar mass as measured by \textsc{SINOPSIS}. We define the star formation rate per unit of stellar mass, i.e. the specific star formation rate ($sSFR =SFR/M_\ast$), and then consider as star-forming the galaxies with 
$sSFR>10^{-12}yr^{-1}$. The remaining galaxies are taken as  passive. 
The second definition is based on galaxy colours as measured by LePhare. To identify the threshold in colour that best separates the blue and red populations, we investigate the correlation between sSFR, $\rm (g-r)_{rest-frame}$ colour and $M_r$, as shown in Fig. \ref{cmd_abs_g-r_LePh}, for the subsample analysed by both LePhare and \textsc{SINOPSIS}. 
Passive galaxies, shown with red points, are mostly clustered at $\rm (g-r)_{rest-frame} \geq 0.6$, while star-forming galaxies, colour-coded according to their sSFR, show a broader distribution. Galaxies having $\log(sSFR/yr^{-1})>$-9.8 most likely have $\rm (g-r)_{rest-frame} < 0.6$, while galaxies with redder colours have on average $\log(sSFR/yr^{-1})\sim-10$. We therefore consider  galaxies with $\rm (g-r)_{rest-frame} > 0.6$ to be  `red', and the rest  `blue'.
With this cut, 80\% of passive galaxies are located in the red region of the diagram.

\subsection{Dependence of the galaxy fractions on environment}

\begin{figure*}
\begin{center}
\includegraphics[scale=0.5]{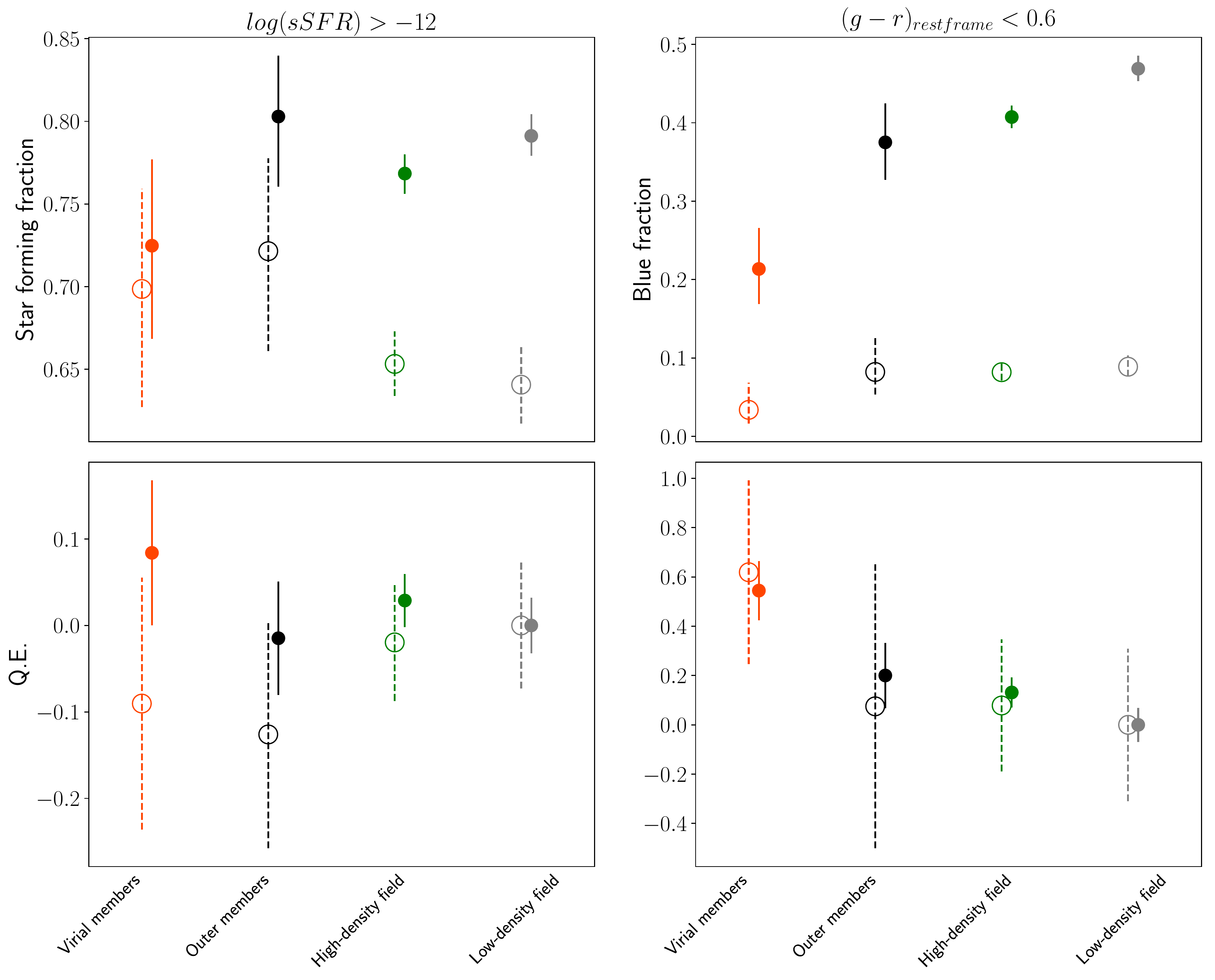}
\caption{Fraction of star-forming galaxies in different environments, computed with sSFR (left panel) and rest-frame colour (right panel). The fractions obtained using the magnitude-limited sample are represented with filled symbols and solid errors, those  obtained using the mass-limited sample are represented by empty symbols and dashed error bars. Errors are derived by means of a bootstrap method. The two lower panels show the quenching efficiency (Q.E.) in different environments, computed with equation \ref{QE_eq} for both the star-forming and blue samples. The Q.E. of field galaxies, which is by definition set to zero (see Eq. \ref{QE_eq}), is shown in both panels as a reference. The error bars on the Q.E of the low-density field depend on the amplitude of the confidence intervals associated with the fractions of star-forming and passive galaxies from bootstrapping.}
\label{SFing_frac_plot}
\end{center}
\end{figure*}

We are now in the position of computing the fraction of blue galaxies and star-forming galaxies  separately in the different environments identified in the XLSSsC N01 region (see 
Figure \ref{SFing_frac_plot}).
Focusing on the star-forming fractions  in the magnitude-limited sample (top left panel,  filled symbols), the fraction of star-forming galaxies in virial members is $0.72^{+0.06}_{-0.05}$, whereas that in low-density field galaxies is $0.79^{+0.01}_{-0.01}$; galaxies in the high-density field have an intermediate star-forming fraction ($0.77^{+0.01}_{-0.01}$). The fraction of star-forming galaxies in outer members is slightly higher with respect to the other environments ($0.80^{+0.04}_{-0.04}$) and in particular with respect to virial members.

Similar trends are visible when the rest-frame colour is considered (top right panel of Figure \ref{SFing_frac_plot}), where the reduced size of error bars with respect to the star-forming fraction panel at the top left further confirms the results: in the magnitude-limited sample, the fraction of blue galaxies among the virial members is significantly lower than that in the other environments, being only $0.21^{+0.05}_{-0.04}$. By contrast, outer members and high-density field have similar values within the error bars,  $0.38^{+0.05}_{-0.05}$ in the former and $0.41^{+0.01}_{-0.01}$ in the latter, and galaxies in the low-density field represent $0.47^{+0.02}_{-0.02}$ of the entire sample. The decrease in the fraction of blue galaxies from the low-density field to the virial members population is a factor of $\sim$2.3 and from outer to virial members a factor of $\sim$1.8.

In the mass-limited sample (shown with empty symbols and dashed error bars) the fractions decrease in all environments and the differences between different environments are smoothed. The only difference that is maintained is between the fraction of blue galaxies in virial members of clusters and the blue fraction in the field. The enhancement and subsequent decrease in the fraction of star-forming/blue galaxies going from the field to outer and then virial members, both in the magnitude and in the mass-complete regimes, point to the direction of an environmental effect that influences the evolution of these galaxies. 

In the mass-limited sample, any possible trend is washed out because our stellar mass limit at the redshift of the supercluster selects only high-mass galaxies ($\rm log(M_{\ast}/M_{\odot}) \geq 10.8$) whose star formation activity, according to the downsizing scenario, was concentrated at earlier epochs and on shorter timescales before the onset of mass quenching. The fraction of star-forming/blue galaxies in the mass-limited sample is indeed lower than the corresponding fraction in the magnitude-limited sample in all environments.

We also note that the fractions of blue and star-forming galaxies are different.
In addition to the differences due to the different methods in which the two characteristics are derived, i.e. the star formation rate from spectroscopy and galaxy colours from photometry, the two definitions of star-forming (sSFR) and blue (rest-frame colour) have different physical meanings. Indeed, while the SFR is a snapshot measuring the number of stars produced by the galaxy at the moment it is observed, the colour is also sensitive  to the past history of the galaxy itself, especially the recent history, being determined by its predominant stellar population.
Furthermore, colour is also influenced by other phenomena, such as metallicity and dust extinction.

Following \cite{Nantais2017}, we define the quenching-efficiency parameter (Q.E.) of a given environment with respect to the low-density field as
\begin{equation}
Q.E. = \frac{F_{passive/red,i} - F_{passive/red,low-density field}}{F_{star-forming/blue,low-density field}}
\label{QE_eq}
,\end{equation}
where $F_{passive/red,i}$ is the fraction of passive/red galaxies in that environment, $F_{passive/red,low-density field}$ is the fraction of passive/red galaxies in the low-density field, and $F_{star-forming/blue,low-density field}$ is the fraction of star-forming/blue galaxies in the low-density field.
Values of Q.E. in the different environments for the different subsamples are shown in the lower panels of Figure \ref{SFing_frac_plot}.
In both panels, we show as reference the low-density field, which by definition has Q.E.=0 (Eq. \ref{QE_eq}). We note that the error bars  associated with the low-density field Q.E. are related to the uncertainties on the star-forming/blue and passive/red fractions, i.e. on the amplitude of the confidence intervals computed through the bootstrapping method.

The efficiency with which cluster virial regions suppress star formation stands out: the Q.E. is significantly higher than in the other environments. The trend is particularly significant using the colour fractions:
the Q.E is $\sim$ 0.5 in cluster virial members, decreases to $\sim$0.2 in outer members, and to $\sim$0.1 in the high-density field. 

Values computed using the sSFR fractions are lower: the Q.E. is close to zero in all environments except as traced by virial members, where it reaches a value of $\sim$ 0.1.  In the magnitude and the mass-limited samples, outer members are characterised by a negative Q.E. (decreasing from $\sim$-0.01 to $\sim$-0.13 going from one sample to another), suggesting an enhanced star formation activity.
In the mass-limited sample (bottom left panel of Figure \ref{SFing_frac_plot}), however, the negative value of the Q.E. of virial and outer members simply reflects the corresponding star-forming fractions above, and in both cases the large error bars prevent us from drawing solid conclusions.

\subsection{The sSFR-- and SFR--Mass relations in different environments}
\begin{figure*}
\begin{center}
\includegraphics[scale=0.425]{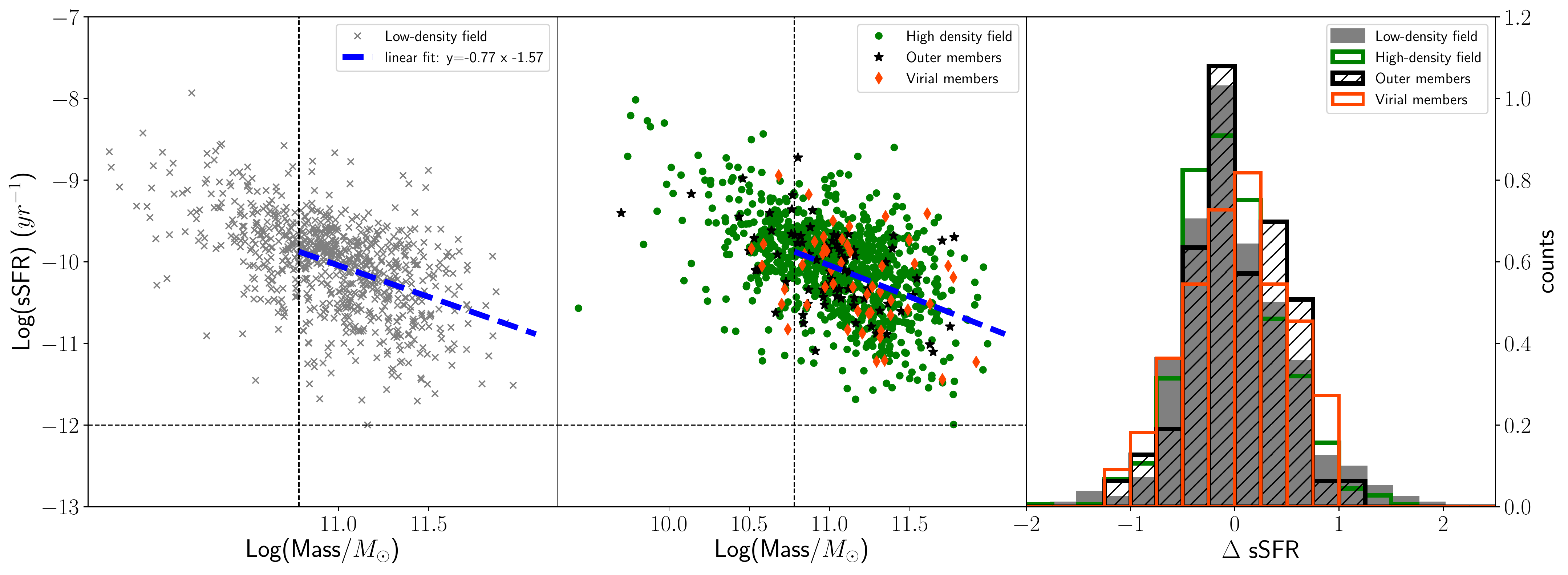}
\caption{Specific star formation rate (sSFR)--mass relation for galaxies in the low-density field (left panel), and galaxies in the high-density field and cluster virial and outer regions (green dots, orange diamonds, and black stars in the central panel). The vertical and horizontal lines show the stellar mass limit and our adopted separation between star-forming and passive galaxies. The blue dashed line is the fit to the relation of the sample including all the environments.
The right panel shows the distribution of the differences between the galaxy sSFRs and their expected values according to the fit given their mass. 
}
\label{ssfr_mass_rel}
\end{center}
\end{figure*}

In  the previous subsection we  detected a dependence of the star-forming and blue fractions on environment.
We here correlate the galaxy star-forming properties with the stellar mass to further inspect the role of the environment.

First, we focus on the sSFR. In Fig.~\ref{ssfr_mass_rel} we show  the sSFR--mass relation in the four environments introduced above.
Very little difference is observed between the different samples, at least above the mass completeness limit. 
To better quantify the differences, we compute the linear regression fit taking together the different environments above the mass completeness limit. We then plot the distribution of the difference between the  sSFR of each galaxy and the value derived from the fit (right panel of Fig. \ref{ssfr_mass_rel}). 
We perform Kolmogorov--Smirnov statistical tests (KS) to compare these distributions, and find that they are all compatible with being drawn from a single parent sample (i.e. the p-values are above the significance level of 0.05).
Overall, we conclude that the sSFR--Mass relation does not seem  to depend on global environment above the galaxy stellar mass limit in our sample, even considering extremely different environments such as X-ray clusters within the XLSSsC N01 supercluster and the field (uncontaminated by X-ray groups or clusters).

\begin{figure}
\begin{center}
\includegraphics[scale=0.6]{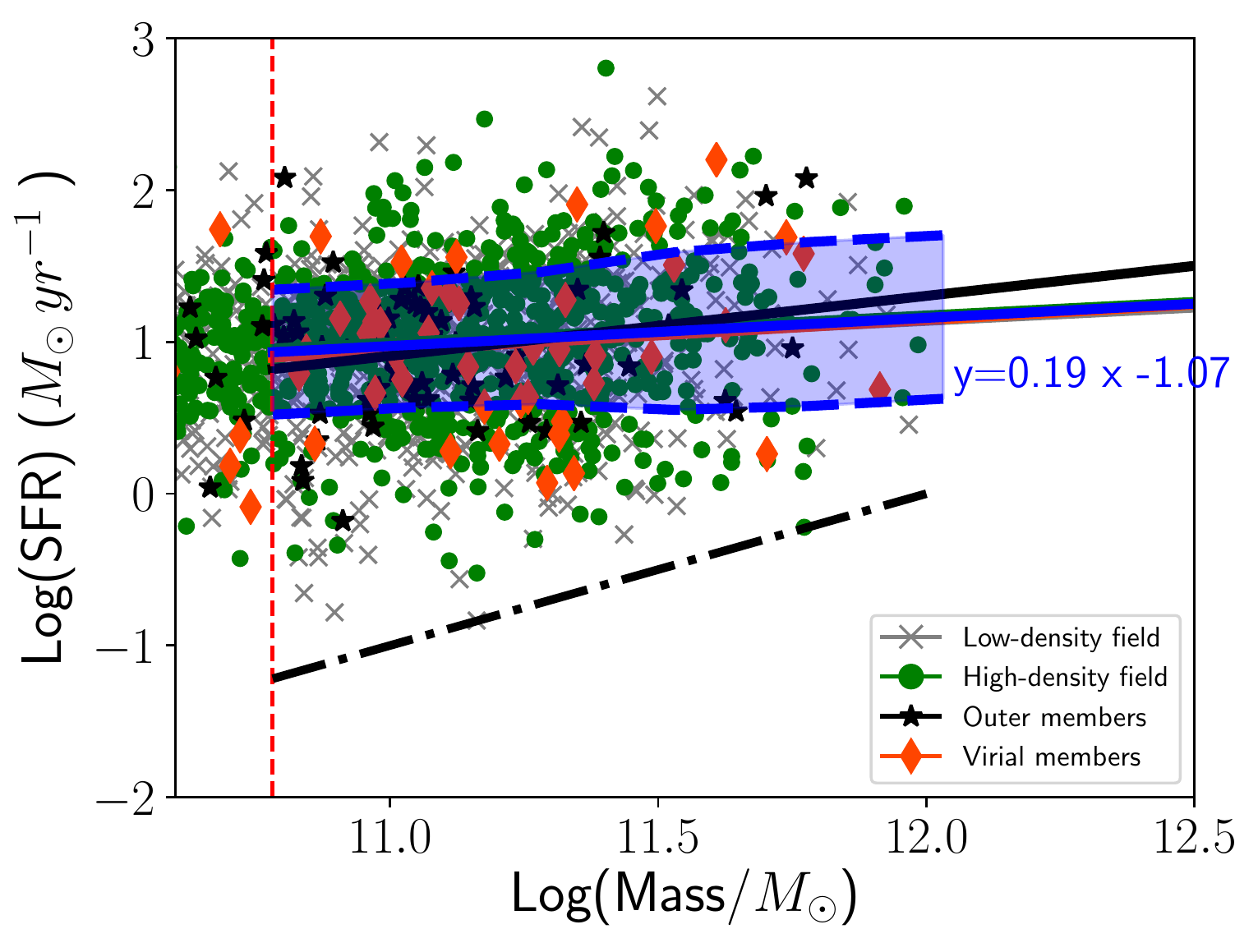}
\caption{SFR--mass relation for galaxies in the low-density field (grey crosses),  in the high-density field (green dots), cluster virial (orange diamonds), and outer members (black stars). The red dashed vertical line shows the stellar mass limit. The blue line is the fit to the relation including all the environments, and the shaded areas correspond to 1$\sigma$ errors on the fitting line. Linear fits for each environment are shown separately in the figure, colour-coded according to the legend.
The black dashed line represents the log(sSFR)=-12 limit.}
\label{sfr_mass_tr}
\end{center}
\end{figure}

Second, we focus on the SFR--Mass relation of star-forming galaxies in the four defined environments, shown in Fig. \ref{sfr_mass_tr}. It is evident that galaxies in the different environments are similarly distributed in this plane. In fact, the linear fits performed on the data points of each environment separately are completely superposed when considering virial members and high- and low- density fields, and is slightly steeper when the population of  outer members  is considered, although the difference can be appreciated only at masses higher  than those covered by our sample. We perform a sigma-clipping linear fit to the relation in the mass-complete regime, and compute 1$\sigma$ confidence intervals, which are shown as blue shaded areas around the solid blue fitting line. Following \citet{Paccagnella2016}, we identify galaxies in transition between the star-forming main sequence and the quenched population as those galaxies with $log(sSFR/yr^{-1})>-12$ and SFR below -1$\sigma$ with respect to the SFR-mass fitting line. The fraction is computed as the ratio of this population to the population of galaxies with $log(sSFR/yr^{-1})>-12$, in each environment.
We find that the incidence of galaxies in transition is not environment dependent, being 0.19$^{+0.03}_{-0.02}$ in the low-density field, 0.16$^{+0.02}_{-0.02}$ in the high-density field, 0.19$^{+0.86}_{-0.68}$ in cluster outskirts, and  0.18$^{+0.10}_{-0.08}$ in the virial regions of clusters. This trend suggests that in the regions surrounding the XLSSsC N01 supercluster the migration from the star-forming main sequence to the quenched stage occurs similarly from the innermost regions of clusters, to the outskirts, and to the surrounding field.

\subsection{Luminosity-weighted age in different environments}

\begin{figure}
\includegraphics[scale=0.45]{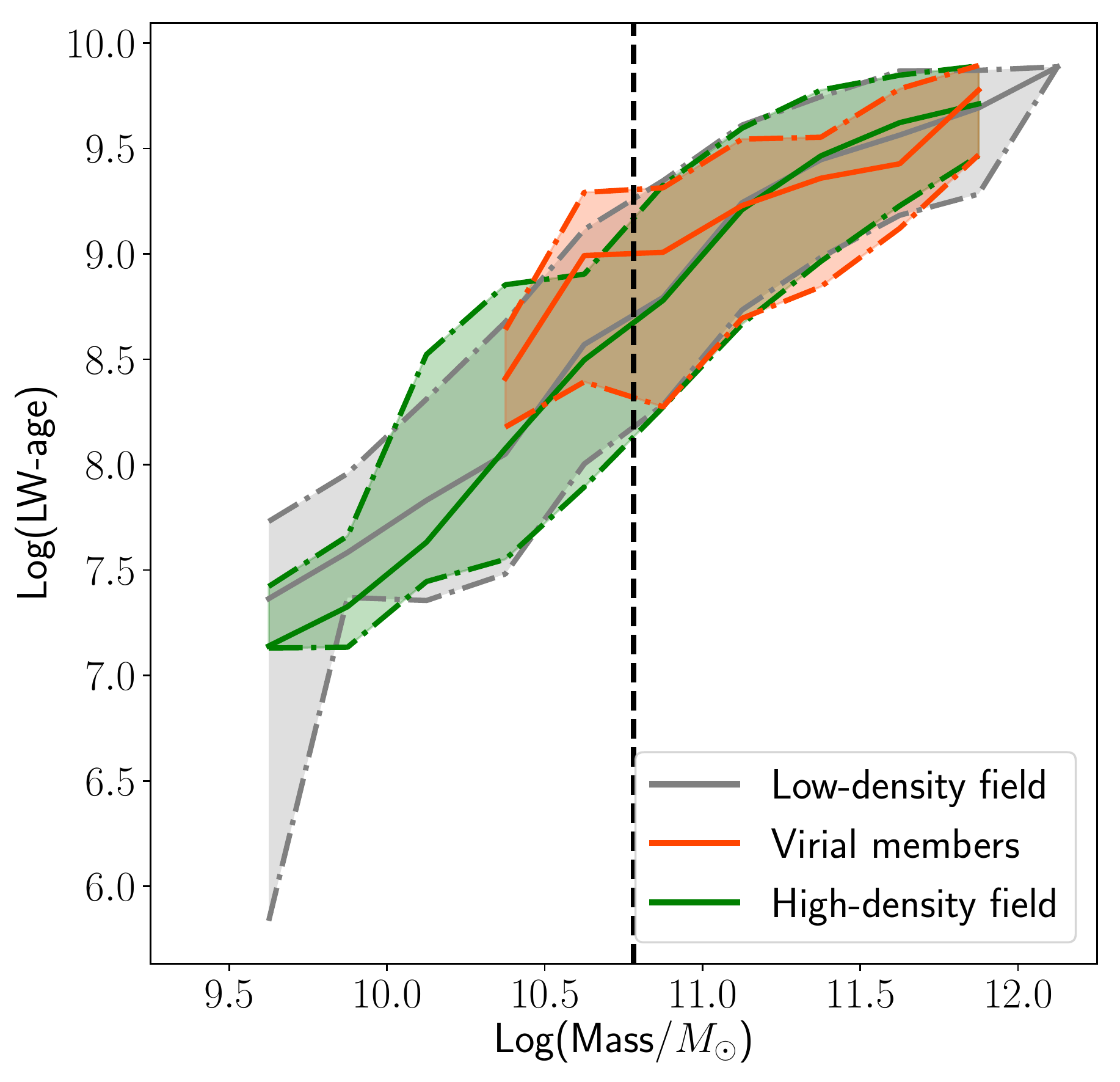}
\caption{Median luminosity-weighted age--mass relation computed in non-independent stellar mass bins for different environments, as shown in the legend. The stellar mass limit is shown with a vertical black dashed line. Shaded areas are the 32nd and 68th percentiles, corresponding to 1$\sigma$ error bars.}
\label{lwage_mass}
\end{figure}

\begin{figure*}
\begin{center}
\includegraphics[scale=0.45]{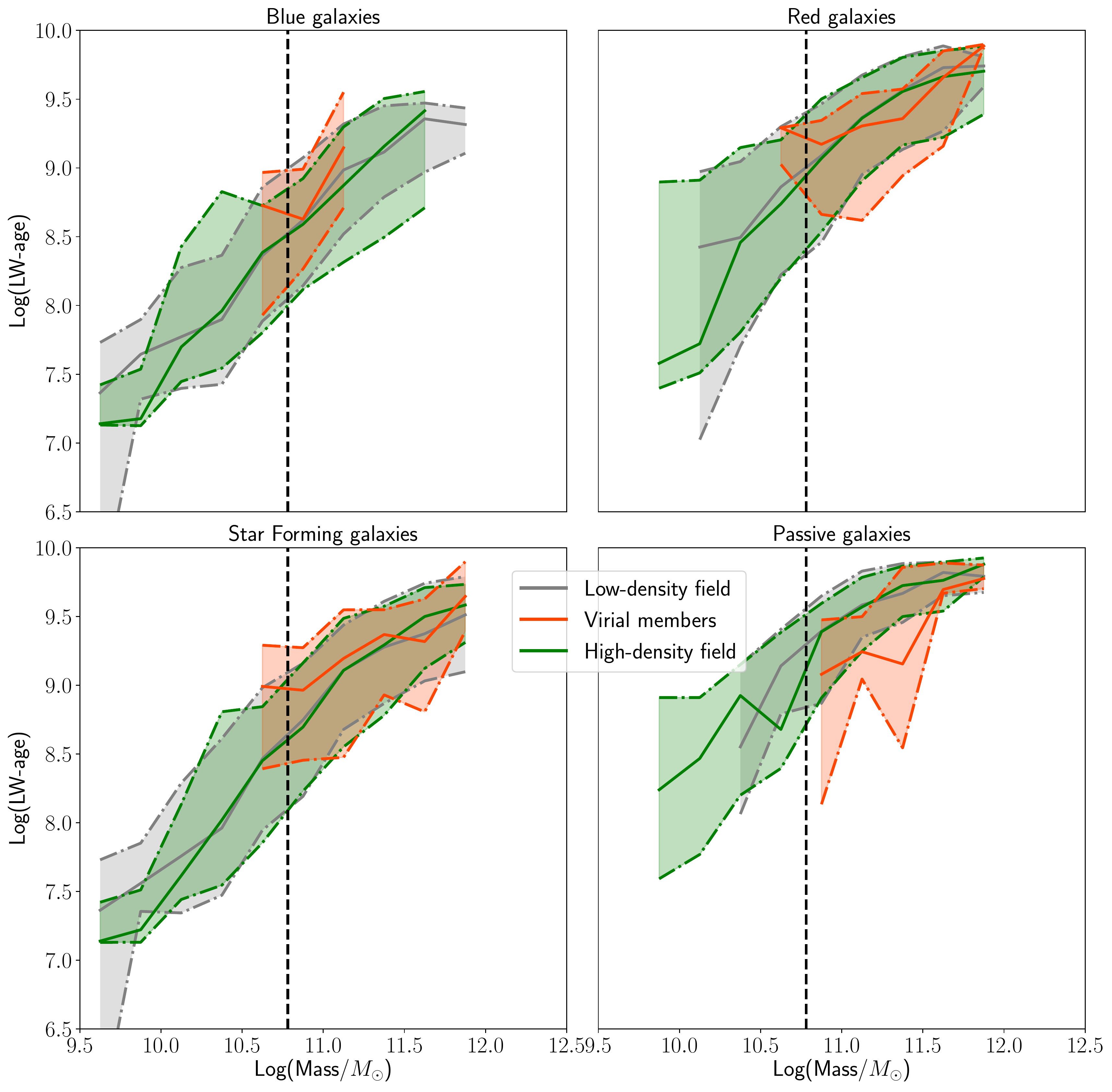}
\caption{Median luminosity-weighted age--mass relation computed in non-independent stellar mass bins for different environments, as shown in the legend, for star-forming/blue and passive/red galaxies. The stellar mass limit is shown with a vertical black dashed line. Shaded areas are the 32nd and 68th percentiles, corresponding to 1$\sigma$ error bars.}
\label{lwage_mass_sf}
\end{center}
\end{figure*}

Differently from the current SFR and sSFR that give information on the ongoing efficiency of galaxies of producing stars, the luminosity-weighted age (LW-age) provides an estimate of the average age of the stars weighted by the light we observe, and it largely reflects the epoch of the last star formation episode.

Figure \ref{lwage_mass} contrasts the median LW-age--mass relation in cluster virial members and in high- and low-density fields in the magnitude-limited sample (the galaxy stellar mass limit is shown with a vertical black dashed line). Medians are computed in non-independent stellar mass bins (i.e. there is an overlapping regime between any given stellar mass bin and the two adjacent ones) in order to minimise statistical fluctuations and shaded intervals correspond to the 32nd and 68th percentiles, which is  the 1$\sigma$ confidence interval. The mean LW-age values span the range from $\sim 1.8 \times 10^{7} yr$ to $\sim 5.6 \times 10^{9} yr$ in the magnitude-limited sample, and from $\sim 10^{9} yr$ to $\sim 5.6\times 10^{9} yr$ in the mass-limited sample. We find overall that the LW-age increases with the galaxy stellar mass in an environmentally independent fashion. No dependences are found even when the outer member population is considered (the results for this population are not shown in the plot for clarity).

To evaluate any possible dependence on galaxy populations, we split galaxies into star-forming/blue and passive/red using the same criteria adopted in previous sections and we plot their median LW-age in the four panels of Figure \ref{lwage_mass_sf}.
While neither the blue nor the red population shows  variations in the median value of the LW-age with environment at any stellar mass, environmental dependences are visible when we use the sSFR to separate the star-forming and passive populations.
In fact, considering the passive populations, the LW-age of passive galaxies in the virial regions of clusters is systematically lower than that of all other galaxies having the same stellar mass, indicating that these galaxies underwent a recent quenching of the star formation, most likely upon accretion into the cluster.

\section{Summary and discussion}
In this work we have presented the characterisation of one of the superclusters identified in XXL Paper XX by means of a FoF algorithm on XXL X-ray clusters. The supercluster, named XLSSsC N01,  has a mean redshift of 0.2956,  is composed of 14 clusters covering a region of 37 Mpc in RA $\times$ 50 Mpc in Dec. Within this region, we characterised the environment of galaxies in 11 clusters, distinguishing among cluster virial members, cluster outer members, high- and low-density field galaxies (which were defined based on their local density).
We  then characterised the properties of the stellar populations of galaxies in these  environments.

First we computed the fraction of star-forming galaxies, defined in terms of their sSFR, and the fraction of blue galaxies, defined in terms of their $(g-r)_{rest-frame}$ colour. 
We note that these definitions have different physical meanings. The SFR is a snapshot measuring the number of stars being produced by the galaxy at the moment it is observed, the colour is also sensitive  to the past history of the galaxy itself, especially the recent history.

The effect of the environment is mainly visible in galaxies in the densest environments (cluster virial/outer members). Indeed, in the magnitude-limited sample ($r\leq 20.0$), the fractions of star-forming and blue galaxies are systematically lower among virial members than in the other environments, indicating star formation quenching in the cluster virialised regions. By contrast, there are hints that the fraction of star-forming galaxies is enhanced among the population of outer members, even with respect to the high- and low-density field. Even though error bars prevent us from stating this on secure statistical grounds, this result suggests an enhancement of star formation when galaxies approach the cluster outskirts.
This result is supported by several works at similar and higher redshifts finding that a number of cluster galaxies mostly located in the outskirts and infalling cluster regions still have large amounts of gas, and their star formation can be triggered by the interactions with the ICM, with other galaxies, or by tidal effects \citep{Bai2007,Marcillac2007,Fadda2008,Santos2013}.
\cite{Marcillac2007} have found that the mid-infrared selected galaxies in a distant cluster (z$\sim$0.83) are associated with infalling galaxies. \cite{Bai2007} have suggested that the cluster environment is able to stimulate the star formation activity in infalling field galaxies before they enter the cluster central regions where gas is stripped and star formation subsequently suppressed.
Interestingly, in \citet[][XXL Paper XXXV]{Koulouridis2018} the  same behaviour was found for active galactic nuclei (AGNs) in the XXL clusters, i.e. AGN activity is enhanced in the outskirts while it steeply drops towards the cluster centres.
In a redshift range similar to that explored in our work, \cite{Fadda2008} found two filamentary structures in the outskirts of the cluster Abell 1763 (z$\sim$0.23), corresponding to infalling galaxies and galaxy groups. The star formation is clearly enhanced in galaxies along the filaments as their associated fraction of starburst galaxies is more than twice  that in other cluster regions. They speculate that the relatively high density of galaxies in filaments compared to the general field, and their relatively low velocity dispersion enhances the tidal effect of galaxy encounters and hence the probability of an induced star-forming activity.
An enhancement in star formation of galaxies located in the cluster outskirts was also found  at high redshift (z$\sim$1.4) by \cite{Santos2013}, which associated most of the measured FIR star formation in a massive distant cluster with potentially infalling galaxies at the edge of the cluster X-ray emission.

Furthermore, when considering the colour fractions, some additional environmental effects might emerge. In addition to an enhancement of the blue population in the outer members with respect to virial members, the high-density field behaves similarly to the outer members and has a lower incidence of star-forming galaxies than the low-density field. Overall, considering the colours, there is a monotonic trend of increasing star-forming fractions from the clusters, to the high-density and to the low-density field.
We note that the less pronounced enhancement in the blue fraction of outer members compared to their star-forming counterpart may be caused by the fact that the environmental mechanisms triggering star formation are not strong enough to change the colour of these galaxies. Conversely, the star formation enhancement is noticeable when measuring the star formation activity directly from emission lines.

These findings are validated by the quenching efficiency parameter that was computed in all environments using the low-density field as reference.
In the local Universe, similar conclusions were drawn by \cite{Wetzel2012}, who detected a significant quenching enhancement around massive clusters only for galaxies closer than 2 virial radii from the centre.

In the mass-limited sample ($\log M_\ast/M_\odot \geq 10.8$), the fractions of  star-forming and blue galaxies are both lower than their corresponding value in the magnitude-limited sample, quenching efficiency trends are flatter and  differences between environments are no longer evident. The stellar mass limit selects only the high-mass end of the galaxy stellar mass function, thus the absence of  environmental dependences suggests that the evolution of massive galaxies is mostly completed by this epoch, in agreement with what was previously found (e.g. \citet{Brinchmann2004,Peng2010,Woo2013}).
This scenario is consistent with the downsizing effect \citep{Cowie1996}, according to which galaxies with higher masses are on average characterised by shorter and earlier star formation processes, and become passive on shorter timescales than lower mass galaxies.

We have also investigated the sSFR-mass relation, and find no difference among galaxies in the low- and high- density fields, and in clusters.
These results differ from previous findings at similar redshift where a population of cluster galaxies were identified with reduced sSFR with respect to the field at any given stellar mass \cite[e.g.][]{Vulcani2010,Patel2009}.

The differences might be primarily due to the fact that here we are investigating low-mass clusters and groups, while previous works studied more massive structures. \citet{Vulcani2010}, for example,  found no differences between the sSFR-mass relation of groups and the field.
Furthermore, the fact that the sSFR-mass relation does not show any dependence on environment, while the fraction of star-forming galaxies does, points towards fast quenching mechanisms leading to the formation of a passive population without any evidence of transition in the sSFR-mass diagram. 
Indeed, the fraction of galaxies in transition from being star-forming to passive, being below 1$\sigma$ of the SFR--mass relation, is similar throughout the different environments explored in the region surrounding XLSSsC N01.

Finally, we have explored the LW-age--mass relation, finding a systematic increase in the mean LW-age with increasing stellar mass, once again in agreement with the downsizing scenario.
Furthermore, while the median LW-age--mass relation of the global population of galaxies is independent of environment, a clear signature of recent quenching of the star formation activity emerges in the passive population of galaxies in the virial regions of X-ray clusters, suggesting the action of environmental processes which are also responsible for the drop in the star-forming fractions highlighted above.

As one of the first studies on stellar populations and star formation activity within superstructures, this analysis lays the groundwork for future investigation into the properties of stellar populations of galaxies on larger samples.
In a future work (Guglielmo et al., in prep.), we will investigate whether the characteristics of XLSSsC N01 are shared by other X-ray structures and superstructures by taking advantage of the improved sample statistics of the whole sample of XXL superclusters, with the possibility of exploring a redshift range from $z=0.1$ up to $z=0.5$. This will also allow a direct comparison with X-ray clusters not belonging to superclusters and will shed light on the role of the large-scale structure in galaxy evolution.

\begin{acknowledgements}

XXL is an international project based around an XMM Very Large Programme surveying two 25 deg$^2$ extragalactic fields at a depth of $\sim 5 \times 10^{-15} {\rm erg \, s^{-1} \, cm^{-2}}$ in the [0.5--2] keV band for point-like sources. The XXL website is http://irfu.cea.fr/xxl. Multi-band information and spectroscopic follow-up of the X-ray sources are obtained through a number of survey programmes, summarised at http://xxlmultiwave.pbworks.com/. The Australia Telescope Compact Array is part of the Australia Telescope National Facility, which is funded by the Australian Government for operation as a National Facility managed by CSIRO.
GAMA is a joint European-Australasian project based around a spectroscopic campaign using the Anglo-Australian Telescope. The GAMA input catalogue is based on data taken from the Sloan Digital Sky Survey and the UKIRT Infrared Deep Sky Survey. Complementary imaging of the GAMA regions is being obtained by a number of independent survey programmes including GALEX MIS, VST KiDS, VISTA VIKING, WISE, Herschel-ATLAS, GMRT, and ASKAP providing UV to radio coverage. GAMA is funded by the STFC (UK), the ARC (Australia), the AAO, and the participating institutions. The GAMA website is http://www.gama-survey.org/.
V.G. acknowledges financial support from the Fondazione Ing. Aldo Gini.
B.V. acknowledges the support from an Australian Research Council Discovery Early Career Researcher Award (PD0028506). We acknowledge the financial support from PICS Italy--France scheme (P.I. Angela Iovino).
The Saclay group acknowledges
long-term support from the Centre National d'Etudes Spatiales.
E.K. acknowledges support from the Centre National d'Etudes Spatiales (CNES) and CNRS.
M.E.R.C. and F.P. acknowledge support by the German Aerospace 
Agency (DLR) with funds from the Ministry of Economy and Technology 
(BMWi) through grant 50 OR 1514 and grant 50 OR 1608.
\end{acknowledgements}

\bibliographystyle{aa}
\bibliography{bibliography.bib}

\end{document}